\documentclass[11pt]{article}

\usepackage{amsmath}
\usepackage{amsfonts}
\usepackage{amssymb}
\usepackage{mathrsfs}
\usepackage{color}

\usepackage{epsfig}
\usepackage{dsfont}

\usepackage{multirow}
\usepackage{graphicx}
\graphicspath{{./allimages/}}

\usepackage{makecell} 
\usepackage{boldline}
\usepackage{diagbox}

\usepackage[title]{appendix}

\usepackage{enumerate} 

\usepackage{tikz}
\usetikzlibrary{arrows,chains,matrix,positioning,scopes,automata}
\usepackage{pgfplots}

\usepackage[colorlinks=true]{hyperref}
\hypersetup{urlcolor=blue, citecolor=blue,linkcolor=blue}
\renewcommand{\Box}{\framebox{\rule{0.3em}{0.0em}}}

\newtheorem{theorem}{Theorem}[section]
\newtheorem{theorem*}{Theorem}[subsubsection]
\newtheorem{lemma}{Lemma}[section]
\newtheorem{proposition}{Proposition}[section]
\newtheorem{example}{Example}[section]
\newtheorem{remark}{Remark}[section]
\newtheorem{definition}{Definition}[section]

\newtheorem{corollary}{Corollary}[section]

\newcommand{\setd}{{ d \kern -.15em l}}
\newcommand{\widehatsetd}{ d \widehat{\kern -.15em l }}
\newcommand{\dd}{\mathsf {d\kern -0.07em l}} 

\renewcommand{\Box}{\hfill \rule{2.3mm}{2.3mm}}

\newcommand{\bgeqn}{\begin{eqnarray}}
\newcommand{\edeqn}{\end{eqnarray}}
\newcommand{\bgeq}{\begin{eqnarray*}}
\newcommand{\edeq}{\end{eqnarray*}}
\newcommand{\R}{{\rm I\!R}}

\newcommand{\inmat}[1]{\mbox{\rm {#1}}}

\newcommand{\F}{{\cal F}}

\newcommand{\B}{{\cal B}}

\newcommand{\be}{\begin{equation}}
\newcommand{\ee}{\end{equation}}

\makeatletter
\@addtoreset{equation}{section}
\makeatother

\title{Quantitative Statistical Robustness for Tail-Dependent Law Invariant Risk Measures}

\author{Wei Wang\footnotemark[2]\ \and Huifu Xu\footnotemark[3]\ \and Tiejun Ma\footnotemark[4]
}

\begin{document}

\maketitle

\renewcommand{\thefootnote}{\fnsymbol{footnote}}
\footnotetext[2]{School of Business, University of Southampton,
Southampton, SO17 1BJ, UK (ww1e17@soton.ac.uk).}
\footnotetext[3]{Department of Systems Engineering \& Engineering Management, 
The Chinese University of Hong Kong, Hong Kong (hfxu@se.cuhk.edu.hk), Shatin, N. T., Hong Kong.}
\footnotetext[4]{School of Business, University of Southampton, Southampton, SO17 1BJ, UK (tiejun.ma@soton.ac.uk).}

\renewcommand{\thefootnote}{\arabic{footnote}}

\begin{abstract}
When estimating the risk of a financial position with empirical data or Monte Carlo simulations via a tail-dependent law invariant risk measure such as the Conditional Value-at-Risk (CVaR), it is important to ensure robustness of the statistical estimator particularly when the data contain noise. 
Kr\"atscher et al. \cite{kratschmer2014comparative} propose a new framework to examine the qualitative robustness of estimators for tail-dependent law invariant risk measures on Orlicz spaces, which is a step further from earlier work for studying the robustness of risk measurement procedures by Cont et al. \cite{cont2010robustness}. 
In this paper, we follow the stream of research to propose a quantitative approach for verifying the statistical robustness of tail-dependent law invariant risk measures. 
A distinct feature of our approach is that we use the Fortet-Mourier metric to quantify variation of the true underlying probability measure in the analysis of the discrepancy between the laws of the plug-in estimators of law invariant risk measure based on the true data and perturbed data, which enables us to derive an explicit error bound for the discrepancy when the risk functional is Lipschitz continuous with respect to a class of admissible laws. 
Moreover, the newly introduced notion of Lipschitz continuity allows us to examine the degree of robustness for tail-dependent risk measures. 
Finally, we apply our quantitative approach to some well-known risk measures to illustrate our theory. 
\end{abstract}

\textbf{Keywords.} Quantitative robustness, tail-dependent law invariant risk measures, Fortet-Mourier metric, admissible laws, index of quantitative robustness.


\section{Introduction}

One of the main purposes of quantitative modeling in finance is to quantify the loss of a financial portfolio. 
Over the past two decades, various risk measures have been proposed for measuring the risk of financial portfolios. 
A risk measure is represented as a map assigning an extended real number (a measure of risk) to each random loss under an implicit assumption that the true loss probability distribution is known. 
However, in practice, the true probability distribution is often unknown or it is prohibitively 
expensive to calculate the risk using the true distribution. 
Thus, in applications, 
evaluating the risk of a random variable representing the loss of a financial position 
often involves two steps: 
estimating the probability distribution from available observations or from the sampling data of the random financial loss via, e.g., Monte Carlo method 
and then plugging  the estimated distribution into a risk measure to quantify the financial loss. 
This is because the risk measures are mostly law invariant, 
that is, they 
are determined only by the probability distributions of random variables. 
For the loss of a financial portfolio, a measure of risk computed based on the estimated distribution is known as a \emph{plug-in estimate} for the risk measure \cite{zahle2011rates}.

Let $X$ denote the random loss of a financial portfolio on a probability space $(\Omega,\mathcal{F},\mathbb{P})$ and $\rho$ be a law invariant risk measure. The plug-in estimate for $\rho(X)$ is given by $\varrho(\widehat{P})$, where $\widehat{P}$ is the empirical distribution based on 
available observations 
and $\varrho$ is a \emph{risk functional} defined by \bgeqn
\label{eq:risk-functional-expression}
\varrho(P)=\rho(X),\;\;\inmat{if}\; X\; \inmat{has law}\;P;
\edeqn
see e.g. \cite{belomestny2012central,beutner2010modified}. 
In the literature, 
Cont et al. \cite{cont2010robustness} first study the quality of statistical estimators of the law invariant risk measures 
using Hampel's classical concept of qualitative robustness \cite{hampel1971general}, 
that is, a risk functional estimator is said to be \emph{qualitatively robust} 
if it is insensitive to the variation of the sampling data. 
The research is important because perceived data (particularly empirical data) may contain some noise. 
Without such insensitivity, financial activities based on the risk measures may cause damage. 
For instance, when $\rho(X)$ is applied to allocate the risk capital for an insurance company, altering the capital allocation may be costly. 
According to Hampel's theorem, Cont et al. \cite{cont2010robustness} 
demonstrate that the qualitative robustness of a statistical estimator is equivalent to the weak continuity of the risk functional,
and 
that value at risk (VaR) is qualitatively robust whereas conditional value at risk (CVaR) is not.

Kr\"atschmer et al \cite{kratschmer2012qualitative} argue that the use of Hampel's classical concept of qualitative robustness may be problematic 
because it requires the risk measure essentially to be insensitive with respect to the tail behaviour of the random variable 
and the recent financial crisis shows that a faulty estimate of tail behaviour can lead to a drastic underestimation of the risk. 
Consequently, they propose a refined notion of qualitative robustness that applies also to tail-dependent statistical functionals and that allows us to compare statistical functionals in regards to their degree of robustness. 
The new concept captures the trade-off between robustness and sensitivity and can be quantified by an index of qualitative robustness. 
Furthermore, under the new concept, Kr\"atschmer et al \cite{kratschmer2014comparative} analyze the qualitative robustness to the law-invariant convex risk measure on Orlicz spaces and show that CVaR and spectral risk measures are all qualitatively robust 
when the perturbation of probability distribution is restricted to a finer topological space. 
Alternative generalizations of Hampel's theorem can be found for strong mixing data (Z\"ahle \cite{zahle2014qualitative,zahle2015qualitative}) and for stochastic processes in various ways (Boente et al \cite{boente1987qualitative} and Strohriegl and Hable \cite{strohriegl2016qualitative}). 
For comprehensive study of statistical robustness, 
we refer readers to \cite{ huber2011robust,zahle2016definition,hampel2011robust,maronna2019robust} and references therein.

In this paper, we take a step further by deriving an error bound for the plug-in estimators of law invariant risk measures
in terms of the variation of data and we call the analysis quantitative because no such error bound is established in the existing qualitative robust analysis. This is achieved by adopting different metrics to  measure the discrepancy of the estimators and the 
variation of data.
Specifically, we  use the Fortet-Mourier metrics as opposed to the L\'evy distance  in Cont et al. \cite{cont2010robustness} or the weighted Kolmogorov metric in Kr\"atschmer et al. \cite{kratschmer2012qualitative} to quantify the data variation (the perturbation of the true probability distributions).  
Moreover, we introduce a new notion of the so-called {\em admissible laws}, which effectively restrict the scope of data variation. 
The new metrics enable us to establish an explicit relationship between the discrepancy of the laws of the plug-in estimators (of law invariant risk measure based on the true data and perturbed data) and 
the discrepancy of the associated probability distributions of the data.
The research is inspired by the recent work of Guo and Xu \cite{GuX2020} where the authors derive quantitative statistical robustness for preference robust optimization models under Kantorovich metric.
The main contributions of the paper can be summarized as follows.

First, we introduce the notion of admissible laws induced by a probability metric, which is a class of probability distributions whose discrepancy with the law of the Dirac measure at $0$ is finite. 
The admissibility effectively restricts the scope of data perturbation. Using the notion, we compare the admissibility under $\phi$-topology and the Fortet-Mourier metric.

Second, we propose to use the Fortet-Mourier metric to quantify the variation of the probability measure. 
The metric enables us to establish an explicit relationship between the discrepancy of the laws of the plug-in estimators of law invariant risk measure based on the true data and perturbed data by noise and the change of the true underlying probability measures when the risk functional is Lipschitz continuous on a class of admissible laws. 
We find that the risk functionals associated with the general moment-type convex risk measures are Lipschitz continuous.

Third, we introduce the concept of Lipschitz continuity for a general statistical functional on a class of admissible laws induced by the Fortet-Morier metric and find that for the Lipschitz continuous risk measure, the parameter of the Fortet-Mourier metric allows us to compare the tail-dependent risk measures with regard to their degree of robustness, i.e., the index of statistical robustness.

Fourth, we apply the new approach to examine the  quantitative statistical robustness of a range of well known risk measures, including CVaR, optimized certainty equivalent, shortfall risk measure 
and conclude that under mild conditions, they are all quantitatively robust, and the indexes of quantitative robustness to them are also calculated.

The rest of the paper is organized as follows. 
In Section 2, we set up the background of the problem for research.
In Section 3, we introduce the concept of  Fortet-Mourier metric and admissible laws. 
In section 4, we establish the quantitative statistical robustness theory and compare with the qualitative statistical robustness theory. 
In section 5, we apply our theory to risk measures and give some examples. 
Some technical details are given in the appendix.




\section{Problem statement}

In this section, we discuss the background of statistical robustness in the context of law invariant risk measures.
We begin by a brief review of law invariant risk measures and its estimation, and then move to explain the issues when the data may contain noise.

Let $(\Omega,\mathcal{F},\mathbb{P})$ be 
an atomless probability space, 
where $\Omega$ is a sample space with sigma algebra $\mathcal{F}$ and $\mathbb{P}$ is a probability measure. 
Let $X: (\Omega,\mathcal{F},\mathbb{P}) \rightarrow \R$ be a financial loss and $F_X(x):=\mathbb{P}(X\leq x)$
be the law or the probability distribution of $X$. 
For $p\geq 1$, let $\mathscr{L}^p(\Omega,\F,\mathbb{P})$ ($\mathscr{L}^p$ for short) denote 
the space of random variables mapping from $(\Omega,\F,\mathbb{P})$ to $\R$ with finite $p$-th order moments. 
We say that a map $\rho: \mathscr{L}^1 \to \overline{\R}:=\R\cup \{+\infty\}$ is a \emph{convex risk measure}\footnote{We note that the canonical model space for law invariant convex risk measure is $\mathscr{L}^1$ \cite{filipovic2012canonical}.} \cite{follmer2002convex} if it satisfies the following properties: 
\begin{enumerate}[(i)]
    \item \emph{Monotonicity}: $\rho(X)\leq \rho(Y)$ for $X,Y\in \mathscr{L}^1$ with $X\leq Y$ $\mathbb{P}$-almost surely;
    \item \emph{Translation invariance}: $\rho(X+c)=\rho(X)+c$ for $X\in \mathscr{L}^1$ and $c\in \R$;
    \item \emph{Convexity}: $\rho(\lambda X+(1-\lambda)Y)\leq \lambda \rho(X)+(1-\lambda)\rho(Y)$ for $X,Y\in \mathscr{L}^1$ and $\lambda\in [0,1]$.
\end{enumerate}
Moreover, if $\rho$ satisfies \emph{positive homogeneity}, i.e.,
for any $\alpha \geq 0$, $\rho(\alpha X)= \alpha \rho(X)$, 
then $\rho$ is a \emph{coherent risk measure}, see \cite{artzner1999coherent,follmer2002convex} for the original definitions of these concepts. 
A risk measure $\rho$ is said to be \emph{law invariant} if  $\rho(X)=\rho(Y)$ for $X$ and $Y$ having the same law. We refer readers to F\"ollmer and Weber \cite{follmer2015axiomatic} for a recent overview of risk measures.

As discussed in \cite{cont2010robustness,kratschmer2012qualitative}, 
it is a widely-accepted procedure to estimate the risk of a financial loss by means of a Monte Carlo method or from a set of available observations. 
Such a procedure is particularly sensible when $\rho$ is law invariant. 
The following proposition states that the law invariance of a risk measure $\rho$ is equivalent to the existence of a \emph{risk functional} $\varrho$ in (\ref{eq:risk-functional-expression}).

\begin{proposition} 
\label{P:RM-represent}
Let $\mathscr{P}(\R)$ denote the set of all probability measures on $\R$. 
If $\rho: \mathscr{L}^1\to \overline{\R}$ is a law invariant risk measure, then there exists a unique \emph{risk functional} $\varrho: \mathscr{P}(\R) \to \overline{\R}$ associated with $\rho$ such that for any $X\in \mathscr{L}^1$, 
\bgeqn
\rho(X) = \varrho(\mathbb{P}\circ X^{-1}).
\label{eq:RM-represent}
\edeqn
\end{proposition}
The result is well-known, see for instance Delage et al. \cite{delage2019dice} for random variables defined in $\mathscr{L}^\infty$.
The usefulness of 
the representation is that it naturally captures the law invariance and  
allows one to define any law invariant risk measure directly 
over the space of probability measures $\mathscr{P}(\R)$ induced by random variables 
in $\mathscr{L}^\infty$ (also known as probability distributions), 
see Fritelli et al. \cite{frittelli2014risk}.
Dentcheva and Ruszczy\'nski \cite{dentcheva2014risk} 
take it further to define 
a class of law invariant risk measures in the space of quantile functions directly.
In a more recent development, Haskell et al. \cite{haskell2018preference} extend the research to a broad class of multi-attribute choice functions defined over the space of survival functions. 
Let $P :=\mathbb{P}\circ X^{-1}$ be the push-forward probability measure on $\R$ induced by $X$.  
Since $\mathbb{P}(X\leq x)$
coincides with $P((-\infty,x])$ ($P(x)$ for short), 
we also call $P$ the distribution or the law of $X$ interchangeably throughout the paper. 
Consequently, we can write 
(\ref{eq:RM-represent}) as (\ref{eq:risk-functional-expression}).

In this paper, we are not concerned with the definition of risk measures over the space of probability distributions or the space of quantile functions, rather we concentrate on the stability of statistical estimators of law invariant risk measures. 
The risk functional $\varrho(P)$ with the law $P=\mathbb{P}\circ X^{-1}$ can be used in a natural way to construct an  estimator for the risk $\rho(X)$ of $X\in \mathscr{L}^1$.
All one needs to do is to take an estimate $P_N$ of $P$ based on the available observations of $X$ and then to plug this estimator into the risk functional $\varrho$ to obtain the desired estimator of $\rho(X)$, i.e.,
\bgeqn
\label{eq:definition-estimator}
\widehat{\varrho}_N(\xi^1,\xi^2,\ldots,\xi^N):=\varrho(P_N),
\edeqn
where in this paper, $P_N$ can be seen as the empirical distribution of an independent and identically distributed (i.i.d., for short) sequence $\xi^1,\xi^2,\ldots,\xi^N$ of historical observations or Monte Carlo simulations, i.e., 
\bgeqn
P_N(x):=\frac{1}{N} \sum_{i=1}^N \mathbf{1}_{\xi^i\leq x}, \quad x\in \R. 
\label{eq:emp-prob-P-N}
\edeqn
Here and later on $\mathbf{1}_{A}$ denotes the indicator function of event $A$. 
Indeed, $P_N$ can be a fairly general estimates, for instance, $P_N$ can be a smoothed empirical distribution based on uncensored data or empirical distribution based on censored data, see, e.g., \cite{zahle2011rates} or empirical distribution based on identically distributed dependent data, see, e.g., \cite{zahle2015qualitative}.

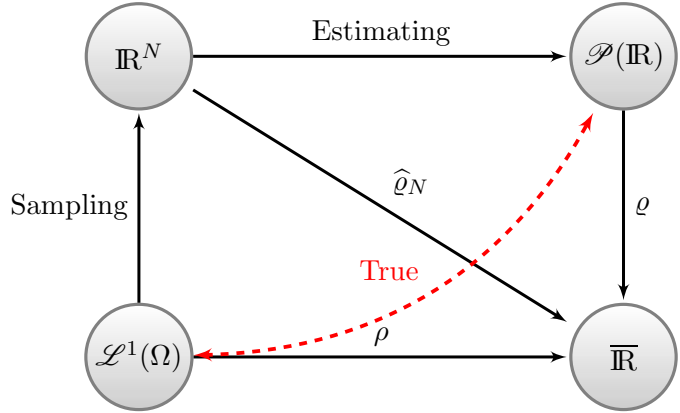
\begin{figure}[hbt!]
\centering
\begin{tikzpicture}[
        > = stealth, 
            shorten > = 1pt, 
            auto,
            node distance = 4cm,
            line width=0.4mm,
            >=latex',scale=.5
        ]
        \tikzstyle{state}=[rectangle,minimum size=14mm,rounded corners=7mm,very thick,draw=black!50, top color=white,bottom color=black!20]
        \node[state] (sample) {$\R^N$};
        \node[state] (omega) [below of= sample] {$\mathscr{L}^1(\Omega)$};
        \node[state] (push) [right = 5cm of sample] {$\mathscr{P}(\R)$};
        \node[state] (r) [below of=push] {$\overline{\R}$};
        \path[->] (sample)  edge node {Estimating} (push);
        \path[->] (omega)   edge node {Sampling} (sample);
        \path[->] (push)    edge node {$\varrho$} (r);
        \path[->] (omega)   edge node {$\rho$} (r);
        \path[->] (sample)  edge node {$\widehat{\varrho}_N$} (r);
        \draw[<->,red, dashed,line width=0.5mm] (omega) to [bend right] node [above left] {True} (push);
\end{tikzpicture}
\caption{\small The diagram for
risk functionals, their estimators and associated spaces}
\label{figure:diagrams}
\end{figure}

We can see that $\widehat{\varrho}_N$ is a mapping from $\R^N$ to $\R$. 
Figure \ref{figure:diagrams} illustrates the relationship between the risk functionals, their estimators and the spaces associated.

In practice, the samples obtained from empirical data may contain noise. 
In that case, we might regard the samples as generated by a perturbed random variable $Y$ with law $Q$, that is, $Q=\mathbb{P}\circ Y^{-1}$. 
Let $\tilde{\xi}^1,\cdots,\tilde{\xi}^N$ be i.i.d samples from $Y$. Then the practical empirical distribution function for estimating the law of $X$ is 
\bgeqn
Q_N(x):=\frac{1}{N} \sum_{i=1}^N\mathbf{1}_{\tilde{\xi}^i\leq  x}, \quad x\in \R,
\label{eq:emp-prob-Q-N}
\edeqn
and the practical estimator is  $\tilde{\varrho}_N=\widehat{\varrho}_N(\tilde{\xi}^1,\cdots,\tilde{\xi}^N):=\varrho(Q_N)$ with perceived empirical data whereas $\widehat{\varrho}_N$ is a statistical estimator with noise being detached. 
Since we are unable to obtain the latter, we tend to use the former as a 
statistical estimator of $\rho(X)$ and this works only if the two estimators are sufficiently close. 

To quantify the closeness, we may look into the discrepancy between the laws of the two estimators under some metric $\dd$, i.e., 
\bgeqn
\label{eq:discrepancy of two estimators}
\dd(\mathrm{law}\{\varrho(P_N)\},\mathrm{law}\{\varrho(Q_N)\})=
\dd\left(P^{\otimes N}\circ \widehat{\varrho}_N^{-1}, Q^{\otimes N}\circ \widehat{\varrho}_N^{-1}\right),
\edeqn
where $P^{\otimes N}$ and $Q^{\otimes N}$ denote the probability measures on measurable space $\left(\R^N,\B(\R)^{\otimes N}\right)$ with marginals $P$ and $Q$ on each $(\R, \B(\R))$ respectively,  $\mathcal B(\R)$ denotes the corresponding Borel sigma algebra of $\R$. 
Since neither $P$ nor $Q$ is known, we want the discrepancy to be uniformly small for all $P$ and $Q$ over a subset of admissible laws on $\mathscr{P}(\R)$ so long as $Q$ is sufficiently close to $P$ under some metric $\dd'$. 
The uniformity may be
interpreted as robustness. 
{\em Qualitative robustness} refers to the case that the relationship between $\dd\left(P^{\otimes N}\circ \widehat{\varrho}_N^{-1}, Q^{\otimes N}\circ \tilde{\varrho}_N^{-1}\right)$ 
and $\dd'(P,Q)$ is implicit
whereas {\em quantitative robustness} refers to the case that the relationship is explicit, i.e., a function of the latter can be used to bound the former, and this is what we aim to achieve in this paper because qualitative robustness
have been well investigated, for instance, in \cite{cont2010robustness,kratschmer2014comparative,kratschmer2012qualitative}.
\section{$\zeta$-metrics and admissible laws}

There are two essential elements in investigating both the qualitative and quantitative statistical robustness of a risk functional: 
One is the specific choice of probability metrics but not just the topologies generated by them, see, e.g., \cite{huber2011robust,cont2010robustness,kratschmer2012qualitative}, to quantify the change of the law $P$ and to estimate the discrepancy between the laws of two estimators, i.e., (\ref{eq:discrepancy of two estimators});  
the other is the determination of the subset $\mathscr{M}$ of admissible laws in $\mathscr{P}(\R)$ (see, e.g., \cite{kratschmer2012qualitative,zahle2015qualitative}), containing all empirical distributions: $\mathscr{M}_{1,\mathrm{emp}}\subset \mathscr{M}$, to restrict the perturbation of the law $P$.
For instance, the subset $\mathscr{M}$ may be specified via some generalized moment conditions, which are interesting in econometric or financial applications. 

To introduce these two essential elements thoroughly, some preliminary notions and results in probability theory and statistics such as $\phi$-weak topology are required. 
We first give a sketch of them to prepare our discussions in the follow-up sections.
%
%
%
%
%
Let $\phi:\R\to [0,\infty)$ be a continuous function and ${\cal M}_1^{\phi} := \left\{P'\in \mathscr{P}(\R): \int_{\R} \phi(t)P'(dt)<\infty\right\}$. 
In the particular case when $\phi(\cdot):=|\cdot|^p$ and $p$ is a positive number, write ${\cal M}_1^p$ for ${\cal M}_1^{|\cdot|^p}$. 
Note that ${\cal M}_1^\phi $ defines a subset of probability measures in $\mathscr{P}(\R)$ which satisfies the generalized moment condition of $\phi$. From the definition, we can see that ${\cal M}_1^{p_2} \subset {\cal M}_1^{p_1}$ for any positive numbers $p_1,p_2$ with $p_1<p_2$ due to H\"older inequality.

\begin{definition}[$\phi$-weak topology]
\label{def-phi-topology}
Let $\phi:\R\to [0,\infty)$ be a \emph{gauge function}, that is, $\phi$ is continuous and $\phi\geq 1$ holds outside a compact set.
Define ${\cal C}_1^\phi$ the linear space of all continuous functions $h:\R\to \R$ for which there exists a positive constant $c$ such that
\[
|h(t)|\leq c(\phi(t)+1), \forall t\in \R.
\]
The \emph{$\phi$-weak topology}, denoted by $\tau_\phi$, is the coarsest topology on
${\cal M}_1^\phi$ for which the mapping $g_h:{\cal M}_1^\phi\to \R$ defined by $g_h(P') :=\int_{\R} h(t) P'(dt),\; \forall h\in {\cal C}_1^\phi$, is continuous. 
A sequence $\{P_l\} \subset {\cal M}_1^\phi$ is said to \emph{converge $\phi$-weakly} to $P\in {\cal M}_1^\phi$ written
${P_l} \xrightarrow[]{\phi} P$ if it converges w.r.t. $\tau_\phi$.
\end{definition}

Clearly, $\phi$-weak topology is finer than the weak topology, and the two topologies coincide if and only if $\phi$ is bounded. 
It is well known (see 
\cite[Lemma 3.4]{kratschmer2012qualitative}) that $\phi$-weak convergence is equivalent to weak convergence, denoted by ${P_l} \xrightarrow[]{w} P$, together with 
$\int_{\R} \phi(t)P_l(dt) \to  \int_{\R} \phi(t)P(dt)$. 
Moreover, it follows by \cite{kratschmer2012qualitative,kratschmer2014comparative} that the $\phi$-weak topology on ${\cal M}_1^\phi$ is generated by the metric
$\dd_\phi:{\cal M}_1^\phi\times {\cal M}_1^\phi\to \R$ defined by
\bgeqn
\dd_\phi(P,Q):=\dd_{\mathrm{Prok}}(P,Q)+\left|\int_{\R} \phi(t)P(dt)-\int_{\R} \phi(t) Q(dt)\right|, 
\label{eq:d-psi}
\edeqn
for $P, Q \in {\cal M}_1^{\phi}$,
where $\dd_{\mathrm{Prok}}: \mathscr{P}(\R)\times \mathscr{P}(\R)\to \R_+$ is the Prokhorov metric defined by  
\bgeqn
\dd_{\mathrm{Prok}}(P,Q):=\inf\{ \epsilon>0: P(A) \leq Q(A^\epsilon)+\epsilon, \forall A \in \mathcal B(\R)\},
\edeqn
where $A^\epsilon:= A + B_\epsilon(0)$ denotes the Minkowski sum of $A$ and the open ball centred at $0$ on $\R$ and $\mathcal B(\R)$ is the corresponding Borel sigma algebra on $\R$. 
We note that the Prokhorov metric metrized the weak topology on $\R$ see, e.g., \cite{gibbs2002choosing}.

\subsection{$\zeta$-metrics}

Instead of exploiting the widely-used probability metrics such as the Prokhorov metric and the weighted Kolmogorov metric in the literature of qualitative robustness \cite{cont2010robustness,kratschmer2012qualitative}, 
we will switch to the so-called metrics with $\zeta$-structure 
to establish the quantitative statistical robustness framework for a risk functional. 
In particular, we will use the well-known Kantorovich metric and Fortet-Mourier metrics. 
The new metrics enable us to establish an explicit relationship between the discrepancy of the laws of the plug-in estimators of law invariant risk measures based on the true data and perturbed data with noise and the discrepancy of the associated true probability measure. 
We begin with a formal definition of $\zeta$-metrics and then clarify the relationships between metrics of $\zeta$-structure and those used in \cite{mizera2010qualitative,cont2010robustness,kratschmer2012qualitative}.

\begin{definition}
Let $P,Q\in \mathscr{P}(\R)$ and $\F$ be a class of measurable functions from $\R$ to $\R$.
The metric with $\zeta$-structure is defined by
\bgeqn
\label{def:zeta-metrics}
\dd_{\mathcal{F}}(P,Q):=\sup_{\psi\in \mathcal{F}}\left|\int_{\R} \psi(\xi)P(d\xi)-\int_{\R} \psi(\xi)Q(d\xi)\right|.
\edeqn
\end{definition}

From the definition, we can see that $\dd_{\mathcal{F}}(P,Q)$
is the maximum difference of the expected values 
of 
the class of measurable functions 
$\mathcal{F}$ with respect to $P$ and $Q$.
$\zeta$-metrics are widely used in the stability analysis of stochastic programming, see 
R\"omisch \cite{romisch2003stability} for an excellent overview.
The specific metrics with $\zeta$-structure that we consider in this paper are
the Kantorovich metric and the Fortet-Mourier metric. The next definition gives a precise description of the two notions.

\begin{definition}[Fortet-Mourier metric]
Let 
\bgeqn
\label{eq:locally lipschitz set}
\mathcal{F}_{p}(\R):=\left\{\psi: \R\rightarrow \R: |\psi(\xi)-\psi(\tilde{\xi})|\leq c_{p}(\xi,\tilde{\xi})|\xi-\tilde{\xi}|, \forall \xi,\tilde{\xi}\in \R\right\},
\edeqn
where $c_{p}(\xi,\tilde{\xi}):=\max\{1,|\xi|,|\tilde{\xi}|\}^{p-1}$ for all $\xi,\tilde{\xi}\in \R$ and $p\geq 1$ describes the growth of the local Lipschitz constants. 
The $p$-th order {\em Fortet-Mourier metric} for $P,Q\in \mathscr{P}(\R)$ is defined by
\bgeqn
\label{def:fortet-mourier-metric}
\dd_{FM,p}(P,Q):=\sup_{\psi\in \mathcal{F}_{p}(\R)}\left|\int_{\R}\psi(\xi)P(d\xi)-\int_{\R}\psi(\xi)Q(d\xi)\right|.
\edeqn
In the case when $p=1$, it is known as the {\em Kantorovich metric} for $P,Q\in \mathscr{P}(\R)$ 
\bgeqn
\label{def:kantorovich-metric}
\dd_{K}(P,Q):= \sup_{\psi\in \mathcal{F}_{1}(\R)}\left|\int_{\R}\psi(\xi)P(d\xi)-\int_{\R}\psi(\xi)Q(d\xi)\right|.
\edeqn
\end{definition}

From the definition, we can see that for any positive numbers $p\geq p'\geq 1$, 
\bgeqn
\label{eq:relation-FM-Kan}
\dd_{FM,p}(P,Q)\geq \dd_{FM,p'}(P,Q)\geq \dd_{K}(P,Q),
\edeqn
which means that $\dd_{FM,p}(P,Q)$ becomes tighter as $p$ increases and they are all tighter than $\dd_{K}(P,Q)$. 
Moreover, the Fortet–Mourier metric metricizes weak convergence on sets of probability measures possessing uniformly a $p$-th moment \cite[p. 350]{pflug2011approximations}. 
Notice that the function $t\rightarrow \frac{1}{p}|t|^p$ for $t\in \R$ belongs to $\mathcal{F}_p(\R)$. 
On $\R$, the Fortet–Mourier metric may be equivalently written as 
\bgeqn
\label{eq:rerepresetation for FM}
\dd_{FM,p}(P,Q)=\int_{\R} \max\{1,|x|^{p-1}\}|P(x)-Q(x)|dx,\;\inmat{for}\;P,Q\in \mathscr{P}(\R),
\edeqn
see, e.g., \cite[p. 93]{rachev1991probability}.

In the next example, we illustrate the relationship between  the existing probability metrics
used in statistical robustness and the metrics with $\zeta$-structure.

\begin{example}
\label{ex:probability-metrics}
A number of well known probability metrics are used in the literature of statistical robustness. 

(i) \emph{The Kantorovich (or Wasserstein) metric}. 
Let $\mathcal{F}_1$ be the set of all Lipschitz continuous functions with modulus being bounded by $1$. 
Then
\bgeqn
\dd_{K}(P,Q):=\int_{-\infty}^{+\infty} |P(x)-Q(x)|dx=\dd_{\mathcal{F}_1}(P,Q).
\label{eq:dfn-Kant-mtc}
\edeqn
Moreover, $\dd_{\mathrm{Prok}}(P,Q)^2\leq \dd_{K}(P,Q)$, see \cite[Theorem 2]{gibbs2002choosing}.

(ii) \emph{The L\'evy distance
} \cite{rachev1991probability}. 
Let $\mathcal{F}$ be the set of functions bounded by 1.
Then 
\[
\dd_{\mathrm{L\acute{e}vy}}(P,Q):=\inf\{\epsilon>0: Q(x-\epsilon)-\epsilon \leq P(x)\leq Q(x+\epsilon)+\epsilon,\;\forall\; x\in\R\} \leq \dd_{\mathcal{F}}(P,Q).
\]
Moreover, $\dd_{\mathrm{L\acute{e}vy}}(P,Q)\leq \dd_{\mathrm{Prok}}(P,Q)$ and $\dd_{\mathrm{L\acute{e}vy}}(P,Q)\leq \dd_{(\phi)}(P,Q)$ for any $\phi\geq 1$, see, e.g., \cite{gibbs2002choosing}. 

(iii) \emph{The weighted Kolmogorov metric} \cite{kratschmer2012qualitative}. 
Let $\phi$ be a \emph{$u$-shaped function}, i.e., a continuous function $\phi:\R\rightarrow [1,+\infty)$ that is non-increasing on $(-\infty,0)$ and non-decreasing on $(0,+\infty)$. Then the weighted Kolmogorov metric is defined as 
\[
\dd_{(\phi)}(P,Q):=\sup_{x\in \R}|P(x)-Q(x)|\phi(x)\leq \dd_{\mathcal{F}}(P,Q),
\]
where $\mathcal{F}$ is the set of all functions bounded by $\phi$. 
Precisely, if $\mathcal{F}$ is the set of all indicator functions $\mathbf{1}_{B}$, 
where $B:=\{(-\infty,\xi],\xi\in \R\}$, then $\dd_{\mathcal{F}}(P,Q)=\dd_{(1)}(P,Q)$, which is known as the \emph{Kolmogorov metric}. Similarly, by letting $\mathcal{F}$ be the set of all weighted indicator functions with weighting $\phi$, one can obtain $\dd_{\mathcal{F}}(P,Q)=\dd_{(\phi)}(P,Q)$.

(iv) \emph{The Prokhorov metric} \cite{kratschmer2012qualitative}.
Let $\mathcal{F}$ be the set of all functions bounded by 1. Then by \cite{gibbs2002choosing},
\[
\dd_{\mathrm{Prok}}(P,Q):=\inf\{ \epsilon>0: P(A) \leq Q(A^\epsilon)+\epsilon, \forall A \in \mathcal B(\R)\} \leq \frac{1}{2}\dd_{\mathcal{F}}(P,Q),
\]
where $A^{\epsilon}:=\{x\in \R: \inf_{y\in A}|x-y|\leq \epsilon\}$. 
Moreover, $\dd_{\mathrm{Prok}}(P,Q)^2\leq \dd_{K}(P,Q)$.

(v) \emph{The Dudley’s (or Bounded) Lipschitz metric}  \cite{mizera2010qualitative}. 
Let $\mathcal{F}_{\mathrm{BL}}$ consist of all Lipschitz continuous $f$ such that $\|f\|_{\infty}+\mathrm{Lip}(f)\leq 1$, where $\|f\|_{\infty}$ denotes the usual sup-norm and $\mathrm{Lip}(f)$ is the Lipschiz constant of the Lipschiz function $f$, then
\bgeq
\dd_{\mathcal{F}}(P,Q)=\sup_{f\in \mathcal{F}_{\mathrm{BL}}}\left|\int f(x) P(dx)-\int f(x)Q(dx)\right|:=\dd_{\mathrm{Lip}}(P,Q).
\edeq
Moreover, $\frac{2}{3}\dd_{\mathrm{Prok}}(P,Q)^2\leq \dd_{\mathrm{Lip}}(P,Q)\leq 2\dd_{\mathrm{Prok}}(P,Q)$, see, e.g., \cite[Section 3]{mizera2010qualitative}. 
\end{example}

\subsection{Admissible laws}

We now turn to discuss another important component in statistical robust analysis,
that is, the subset $\mathscr{M}$ of admissible laws in $\mathscr{P}(\R)$ 
which describes the scope of the perturbation of the law $P$ by a metric. 
This can be motivated by ensuring the finiteness  
of $\dd(P,Q)$. 
To this effect, we formally introduce the concept of admissible laws induced by probability metrics.

\begin{definition}[Admissible laws induced by probability metrics]
\label{def:admissible-laws}
Let $\dd$ be a probability metric on $\mathscr{P}(\R)$. 
The admissible laws induced by $\dd$ are defined as
\bgeqn
\label{eq:def-admissible}
\mathscr{P}_{\dd}(\R):=\{P\in \mathscr{P}(\R): \dd(P,\delta_0)<+\infty\},
\edeqn
where $\delta_0$ denotes the Dirac measure at $0$. 
\end{definition} 
Let $\mathscr{P}_p(\R)$ denote the admissible laws induced by the Fortet-Mourier metrics with parameter $p$ on $\mathscr{P}(\R)$. 
By Definition \ref{def:admissible-laws}, we have  
\bgeqn
\mathscr{P}_p(\R)\notag
&:=& \{P\in\mathscr{P}(\R): \dd_{FM,p}(P,\delta_0)<+\infty\}\\ \notag
&=& \left\{P\in\mathscr{P}(\R):\sup_{\psi\in \mathcal{F}_p(\R)}\left|\int_{\R} \psi(\xi)P(d\xi)-\int_{\R}\psi(\xi)\delta_0(d\xi)\right|<+\infty\right\}\\
&=&\left\{P\in\mathscr{P}(\R):\sup_{\psi\in \mathcal{F}_p(\R)}\left|\int_{\R} \psi(\xi)P(d\xi)-\psi(0)\right|<+\infty\right\}. 
\edeqn
By triangle inequality, this ensures $\dd_{FM,p}(P,Q)<+\infty$ for any $P,Q\in \mathscr{P}_p(\R)$.  

In the following example, we compare the admissible laws induced by different probability metrics.

\begin{example}[Admissible laws induced by probability metrics]
\label{ex:admissible-law-probability-metrics}
We reconsider the admissible laws induced by probability metrics defined in Example \ref{ex:probability-metrics}.

(i) \emph{The admissible laws induced by the Kantorovich (or Wasserstein) metric} are defined as
\bgeq
\mathscr{P}_{K}(\R)
&:=& \{P\in \mathscr{P}(\R): \dd_{K}(P,\delta_0)<+\infty\}\\
&=& \left\{P\in \mathscr{P}(\R): \int_{-\infty}^0 P(x)dx+\int_0^{+\infty} (1-P(x))dx < +\infty \right\}(=\mathscr{P}_1(\R))\\
&=& \left\{P\in \mathscr{P}(\R): \int_{\R}|x| dP(x)< +\infty \right\}\\
&=& \mathcal{M}_1^1,
\edeq
where the second equality follows from the definition of the Kantorovich metric (see, (\ref{eq:dfn-Kant-mtc})). 
To see how the third equality holds, we note that for any $t<0$, we have 
\bgeq
+\infty>\int_{-\infty}^0P(x)dx &=& \int_{t}^0 P(x)dx+\int_{2t}^{t} P(x)dx +\int_{-\infty}^{2t} P(x)dx\\
&\geq& \int_{t}^0 P(x)dx+ \frac{1}{2}P(2t)|2t|
\edeq
Since $-2tP(2t)\geq 0$, then let $t\rightarrow -\infty$, then we have $\lim_{t\rightarrow -\infty} tP(t)=0$. Similarly, we have $\lim_{t\rightarrow +\infty} t(1-P(t))=0$. 
By using integration-by-parts formula (more precisely \cite[Theorem 1.15]{mattila1999geometry}), we obtain the right hand side of the third equality.
The last equality follows from the definition of $\phi$-topology in which case $\phi=|\cdot|$.

(ii) \emph{The admissible laws induced by the L\'evy distance} are defined as 
\bgeq
\mathscr{P}_{\mathrm{L\acute{e}vy}}(\R):=\{P\in \mathscr{P}(\R):\dd_{\mathrm{L\acute{e}vy}}(P,\delta_0)<+\infty\}=\mathscr{P}(\R). 
\edeq
Since $\dd_{\mathrm{L\acute{e}vy}}\leq 1$, then the admissible laws coincide with $\mathscr{P}(\R)$.

(iii) \emph{The admissible laws induced by the weighted Kolmogorov metric} are defined as 
\bgeq
\mathscr{P}_{(\phi)}(\R)
&:=&\{P\in \mathscr{P}(\R):\dd_{(\phi)}(P,\delta_0)<+\infty\}\\
&=&\left\{P\in \mathscr{P}(\R): \sup_{x\leq 0}|P(x)\phi(x)|+\sup_{x>0}|(1-P(x))\phi(x)| <+\infty \right\},
\edeq
which coincides with the set $\mathscr{M}_{1}^{(\phi)}$ defined in Kr\"atschmer at al. \cite[subsection 3.2]{kratschmer2012qualitative}. 

If $\phi$ is bounded on $\R$, then it is straight that $\mathscr{P}_{(\phi)}(\R)=\mathscr{P}(\R)$. 
In the case when $\phi$ is unbounded on $\R$, then
\bgeqn
\label{eq:relation_admissible_law_weighted_K}
\mathcal{M}_{1}^{\phi}\subset \mathscr{P}_{(\phi)}(\R)\subset \bigcap_{\epsilon>0} \mathcal{M}_{1}^{\phi^{1-\epsilon}}.
\edeqn
In what follows, we give a proof for (\ref{eq:relation_admissible_law_weighted_K}). 
Let $P\in \mathcal{M}_1^{\phi}$, since $\phi$ is a $u$-shaped function, then for any $M>0$ and $N\leq 0$, we have 
\bgeq
+\infty>\int_{\R} \phi(x) dP(x) 
&=& \int_0^{+\infty} \phi(x)dP(x)+\int_{-\infty}^0 \phi(x)dP(x)\\
&\geq& \int_0^M \phi(x)dP(x)+\phi(M)(1-P(M))+ \int_N^0 \phi(x)dP(x) + \phi(N)P(N)\\
&\geq & \phi(M)(1-P(M))+\phi(N)P(N),
\edeq
and consequently $\int \phi d\mu\geq \sup_{x\leq 0}|P(x)\phi(x)|+\sup_{x>0}|(1-P(x))\phi(x)|$. 
Thus, $\mathcal{M}_1^{\phi}\subset \mathscr{P}_{(\phi)}(\R)$.

On the other hand, for any $\epsilon>0$, if we let $\phi_{\epsilon}(x):=\phi(x)^{1-\epsilon}$ for $x\in \R$, 
then $\phi_{\epsilon}$ is a gague function. 
Moreover, for any  $P\in \mathscr{P}_{(\phi)}(\R)$,   
there exists a $k<+\infty$ such that 
$k=\sup_{x\leq 0}|P(x)\phi(x)|+\sup_{x>0}|(1-P(x))\phi(x)|$.
To ease the exposition, 
we can assume that the law $P(x)>0$ for any  $x\in \R$. Then
\bgeq
\phi(x)\leq \frac{k}{P(x)}\;\inmat{for}\; x\leq 0\;\;\inmat{and}\; \phi(x)\leq \frac{k}{1-P(x)}\;\inmat{for}\; x> 0.
\edeq
Thus
\bgeq
\int_{\R} \phi_{\epsilon}(x)dP(x)
&=&\int_{-\infty}^{0} \phi_{\epsilon}(x)dP(x)+\int_{0}^{+\infty} \phi_{\epsilon}(x)dP(x) \\
&\leq & k^{1-\epsilon} \int_{-\infty}^{0} \frac{1}{P(x)^{1-\epsilon}}dP(x)+k^{1-\epsilon} \int_{0}^{+\infty} \frac{1}{(1-P(x))^{1-\epsilon}} dP(x)\\
&=& k^{1-\epsilon} \left[\frac{1}{\epsilon}P(x)^{\epsilon}\right]_{-\infty}^0-k^{1-\epsilon} \left[\frac{1}{\epsilon}(1-P(x))^{\epsilon} \right]\\
&=&\frac{1}{\epsilon}k^{1-\epsilon}[P(0)^{\epsilon}-(1-P(0))^{\epsilon}]< +\infty
\edeq
which implies $P\in \mathcal{M}_{1}^{\phi_{\epsilon}}$. Summarizing the discussions above, we obtain (\ref{eq:relation_admissible_law_weighted_K}). 

We note that if $\phi$ is unbounded, then the inclusions in (\ref{eq:relation_admissible_law_weighted_K}) are strict because we can find a counterexample showing equality may fail, see Example \ref{ex:counter-example} in the appendix. 


(iv) \emph{The admissible laws induced by the Prokhorov metric} are defined as 
\bgeq
\mathscr{P}_{\mathrm{Prok}}(\R):=\{P\in\mathscr{P}(\R):\dd_{\mathrm{Prok}}(P,\delta_0)<+\infty\} =\mathscr{P}(\R).
\edeq
Since $\dd_{\mathrm{Prok}}\leq 1$, then the admissible laws coincides with $\mathscr{P}(\R)$.

(v) \emph{The admissible laws induced by the Dudley's (or Bounded) metric} are defined as 
\bgeq
\mathscr{P}_{\mathrm{Lip}}(\R)&:=& \{P\in \mathscr{P}(\R):\dd_{\mathrm{Lip}}(P,\delta_0)<+\infty\}\\
&=&\left\{P\in \mathscr{P}(\R): \sup_{f\in \mathcal{F}_{\mathrm{BL}}} \left|\int_{\R} f(x)P(dx)-f(0) \right| <+\infty \right\}= \mathscr{P}(\R).
\edeq
Since $\dd_{\mathrm{Lip}}\leq 2$, then the admissible laws coincide with $\mathscr{P}(\R)$.
\end{example}

\subsection{Relationship with $\phi$-weak topology}

Since $\phi$-weak topology has been widely used for qualitative robust analysis in the literature whereas we use the topology induced by the Fortet-Mourier metrics for quantitative robust analysis, 
it would therefore be helpful to look into potential connections of the two apparently completely different metrics.
In the next proposition, we look into such connection from
admissible set perspective (which defines the space of probability measures that $P$ is perturbed in both qualitative and quantitative robust analysis),
we find that
$\mathscr{P}_p(\R)$ coincides with ${\cal M}_1^\phi$ for some specific choice of $\phi$ and subsequently show that
the Fortet-Mourier metric is tighter than $\dd_\phi$.

\begin{proposition}
\label{prop:class-FM}
Let  $p\geq 1$ be fixed and 
\bgeq
\phi_p(t):=\left\{
\begin{array}{ll}
      |t|, & \inmat{for}\; |t|\leq 1, \\
      |t|^{p}, & \inmat{otherwise}.
\end{array} 
\right.
\edeq
The following assertions hold.
\begin{itemize}
    \item[(i)] $\mathscr{P}_p(\R)=\mathcal{M}_{1}^{\phi_p}(=\mathcal{M}_1^p)$.
    \item[(ii)]  $\dd_{\phi_p}(P,Q)\leq \sqrt{\dd_{FM,p}(P,Q)}+p\dd_{FM,p}(P,Q),\;\forall P,Q\in \mathscr{P}_p(\R)$.
   \item[(iii)] $\dd_{FM,p}$ metrizes the $\phi_p$-weak topology on $\mathscr{P}_p(\R)$.
\end{itemize}
\end{proposition}

Part (i) of the proposition says that the admissible set $\mathscr{P}_p(\R)$ coincides with 
the set of laws on $\R$ satisfying the generalized moment condition of $\phi_p$.
Part (ii) indicates that $\dd_{FM,p}$ is tighter than $\dd_{\phi_p}$. 
Part (iii) means that the $\phi_p$-weak topology on $\mathscr{P}_p(\R)$ is generated by the metric $\dd_{FM,p}$.

\noindent\textbf{Proof.}
Part (i). Since for any $p\geq 1$, $\frac{1}{p}\phi_p\in \mathcal{F}_p(\R)$, then by the definition of $\mathscr{P}_p(\R)$, we have that $P\in \mathscr{P}_p(\R)$ implies $P\in \mathcal{M}_1^{\phi_p}$ and subsequently, $\mathscr{P}_p(\R)\subset \mathcal{M}_{1}^{\phi_p}$. 

On the other hand, let $P\in \mathcal{M}_{1}^{\phi_p}$, then $\int_{\R} \phi_p(\xi)P(d\xi)<\infty$.
For any $\psi \in \mathcal{F}_{p}(\R)$, we have
\bgeq
|\psi(\xi)-\psi(0)|\leq \max\{1,|\xi|^{p-1}\}|\xi|\leq \max\{|\xi|,|\xi|^p\},\;\inmat{for all}\; \xi\in \R,
\edeq
and consequently,
\bgeq
\left|\int_{\R} \psi(\xi)P(d\xi)-\psi(0)\right| &=& \left|\int_{\R} (\psi(\xi)-\psi(0))P(d\xi)\right|\leq \int_{\R} |\psi(\xi)-\psi(0)|P(d\xi)\\
&=&\int_{\R} \max\{|\xi|,|\xi|^p\}P(d\xi)\leq \int_{\R} \phi_p(\xi)P(d\xi).
\edeq
Therefore, we have
\bgeq
\sup_{\psi\in\mathcal{F}_p(\R)}\left|\int_{\R} \psi(\xi)P(d\xi)-\psi(0) \right|\leq \int_{\R} \phi_p(\xi)P(d\xi)<\infty,
\edeq
and consequently, $\mathcal{M}_1^{\phi_p}\subset \mathscr{P}_p(\R)$.

Part (ii). 
Since $\frac{1}{p}\phi_p\in \mathcal{F}_p(\R)$, then for any $P,Q\in \mathscr{P}_p(\R)$, 
\bgeq
\left|\int_{\R} \phi_p(\xi)P(d\xi)-\int_{\R} \phi_p(\xi)Q(d\xi) \right| \leq p\left|\int_{\R} \frac{1}{p}\phi_p(\xi)P(d\xi)-\int_{\R} \frac{1}{p}\phi_p(\xi)Q(d\xi) \right|
\leq 
p \dd_{FM,p}(P,Q).
\edeq
From 
Example \ref{ex:probability-metrics}(i) and (\ref{eq:relation-FM-Kan}), we have
$\dd_{\mathrm{Prok}}(P,Q)\leq \sqrt{\dd_{FM,p}(P,Q)}$. 
Finally, by the definition of $\dd_{\phi_p}$, i.e., (\ref{eq:d-psi}),  
we obtain the conclusion.

Part (iii) follows straightforwardly from Part (ii). 
\hfill $\Box$

Proposition \ref{prop:class-FM} indicates that despite Fortet-Mourier metric $\dd_{FM,p}$ and  $\dd_{\phi_p}$ are different metrics,
they generate the same topology, 
which confirms the statement at the beginning of this section, i.e., for the qualitative robustness and the quantitative robustness, the specific choice of probability metrics matters but not the topologies generated by them.
To conclude this section, we remark that the subset $\mathscr{M}$ to be used in the definition of qualitative robust analysis will be confined to the set of admissible laws when we adopt the Fortet-Mourier metric for quantitative robust analysis in the next section.

\section{Statistical robustness}

We are now ready to return our discussions to the robustness of statistical estimators of law invariant risk measures that are outlined in Section 2. 

\subsection{Qualitative statistical robustness}

To position our research properly, we begin by a brief overview of the existing results about the qualitative statistical robustness.

\begin{definition}[Qualitative $\mathcal{P}_0$-Robustness \cite{cont2010robustness,kratschmer2014comparative}] 
Let $\mathcal{P}_0$ be a subset of $\mathscr{P}(\R)$ and $P\in \mathcal{P}_0$. 
The sequence $\{\widehat{\varrho}_N\}_{N\in\mathbb{N}}$ of estimators is said to be \emph{qualitatively $\mathcal{P}_0$-robust} at $P$ w.r.t. $(\dd,\dd')$ if for every $\epsilon>0$ there exist $\delta>0$ and $N_0\in \mathbb{N}$ such that for all $Q\in\mathscr{P}_0$ and $N\geq N_0$
\bgeqn
\dd(P,Q)\leq \delta \implies \dd'(P^{\mathbb{N}}\circ \widehat{\varrho}_N^{-1}, Q^{\mathbb{N}}\circ \widehat{\varrho}_N^{-1})\leq \epsilon.
\edeqn
If, in addition, $\{\widehat{\varrho}_N\}_{N\in\mathbb{N}}$ arises as in (\ref{eq:definition-estimator}) from a risk functional $\varrho$, then $\varrho$ is called \emph{qualitatively $\mathscr{P}_0$-robust} at $P$ w.r.t. $(\dd,\dd')$.
\end{definition}

The definition above captures two versions of qualitative statistical robustness proposed by Cont et al. \cite{cont2010robustness} for i.i.d. observations on $\R$ 
with $\dd$ and $\dd'$ being L\'evy distance
and Kr\"atchmer et al. \cite{kratschmer2012qualitative} for i.i.d. observations on $\R$ 
with $\dd=\dd_{(\phi)}$ and $\dd'=\dd_{\mathrm{Prok}}$ respectively. 
Since $\dd_{\mathrm{Prok}}$ is tighter than
$\dd_{\mathrm{L\acute{e}vy}}$, it means 
Kr\"atchmer et al. \cite{kratschmer2012qualitative}
examines the discrepancy of the laws with a tighter metric.
On the other hand, from the definition of $\dd_{(\phi)}$, we can see that it is also tighter than $\dd_{\mathrm{L\acute{e}vy}}$ and 
allows one to capture the difference of distributions at the tail, it means the robust analysis in Kr\"atchmer et al. \cite{kratschmer2012qualitative} is restricted to a smaller class of probability distributions when $Q$ is perturbed from $P$. 
This explains why CVaR is  robust under the criterion of the latter but not the former.


A key result that Kr\"atchmer et al. \cite{kratschmer2012qualitative} establish is the Hampel's theorem which states the equivalence between qualitative statistical robustness and stability/continuity of a risk functional (with respect to perturbation of the probability distribution) under uniform Glivenko-Cantelli (UGC) property of empirical distributions over a specified set.

\begin{definition}[$\mathscr{C}$-Continuity \cite{kratschmer2012qualitative}]
Let $P\in \mathscr{P}(\R)$ and $\mathscr{C}$ be a subset of $\mathscr{P}(\R)$. Then $\varrho$ is called \emph{$\mathscr{C}$-continuous} at $P$ w.r.t. $(\dd,|\cdot|)$ if for every $\epsilon>0$, there exists $\delta>0$ such that for all $Q\in \mathscr{C}$
\bgeq
\dd(P,Q)\leq \delta \implies |\varrho(P)-\varrho(Q)|\leq \epsilon.
\edeq
\end{definition}


\begin{definition}[UGC Property \cite{kratschmer2012qualitative}] 
Let $\mathscr{C}$ be a subset of $\mathscr{P}(\R)$. Then we say that the metric space $(\mathscr{C},\dd)$ has \emph{the UGC property} if for every $\epsilon>0$ and $\delta>0$, there exists $N_0\in \mathbb{N}$ such that for all $P\in \mathscr{C}$ and $N\geq N_0$
\bgeq
P^{\otimes N}\left[(\xi^1,\ldots,\xi^N)\in \R^N: \dd(P,P_N)\geq \delta \}\right]\leq \epsilon.
\edeq
\end{definition}
The UGC property means that convergence in probability of the empirical probability measure to the true marginal distribution uniformly in $\mathscr{C}$ on $\mathscr{P}(\R)$. 
Examples for metrics spaces $(\mathscr{C},\dd)$ having the UGC property can be found in \cite[Section 3]{kratschmer2012qualitative}. 
In particular, it is shown that there exists a subset of the admissible laws induced by the weigthed Kolmogorov metric enjoys the UGC property, see \cite[Theorem 3.1]{kratschmer2012qualitative}.



\begin{theorem}[Hampel's Theorem \cite{kratschmer2012qualitative}] 
Let $\mathscr{P}_0$ be a subset of $\mathscr{P}(\R)$ and $P\in \mathscr{P}_0$. 
Assume that $(\mathscr{P}_0,\dd)$ has the UGC property and $\mathscr{M}_{1,\mathrm{emp}}\subset \mathscr{P}_0$. 
Then if the mapping $\varrho$ is $\mathscr{M}_{1,\mathrm{emp}}$-continuous at $P$ w.r.t. $(\dd,|\cdot|)$, the sequence $\{\widehat{\varrho}_N\}_{N\in\mathbb{N}}$ is qualitatively $\mathscr{P}_0$-robust at $P$ w.r.t. $(\dd,\dd_{\mathrm{Prok}})$.
\end{theorem}

\subsection{Quantitative statistical robustness}

We now move on to discuss our central topic,
quantitative statistical robustness for the plug-in estimators of law invariant risk measures. 
Intuitively speaking, quantitative statistical robustness of a risk functional $\varrho$ means 
that for any two admissible laws $P$ and $Q$ on $\mathscr{P}(\R)$, 
the distance between the laws of their plug-in estimators $\varrho(P_N)$ and $\varrho(Q_N)$
is bounded by the distance between $P$ and $Q$ when the sample size is sufficiently large.

\begin{definition}[Quantitative statistical robustness]
Let $\dd,\dd'$ be probability metrics on $\mathscr{P}(\R)$ and $\mathscr{M}\subset \mathscr{P}_{\dd'}(\R)$ denote a subset of admissible laws on $\R$. 
A sequence of statistical estimators $\{\widehat{\varrho}_N\}_{N\in\mathbb{N}}$ is said to be \emph{quantitative statistical robust} on $\mathscr{M}$ w.r.t. $(\dd, \dd')$ if there exists a non-decreasing real-valued continuous function $h:\R_+\to\R_+$ with $h(0)=0$ such that for all $P, Q \in \mathscr{M}$ and $N\in \mathbb{N}$ 
\bgeqn
\label{eq:def-QSR}
\dd(P^{\otimes N}\circ \widehat{\varrho}_N^{-1},
Q^{\otimes N}\circ \widehat{\varrho}_N^{-1})\leq h(\dd'(P,Q))<+\infty.
\edeqn
If in addition, $\{\widehat{\varrho}_N\}_{N\in\mathbb{N}}$ arise as in (\ref{eq:definition-estimator}) from a risk functional $\varrho$, then $\varrho$ is called \emph{quantitative statistical robust} on $\mathscr{M}$ at $P$ w.r.t. $(\dd,\dd')$.
In a particular case when $\dd=\dd_{K}$, $h(t)=Lt$ and $\dd'=\dd_{FM,p}$, inequality  (\ref{eq:def-QSR}) reduces to
\bgeqn
\label{eq:def-QSR-FM}
\dd_{K}\left(P^{\otimes N}\circ \widehat{\varrho}_N^{-1},
Q^{\otimes N}\circ \widehat{\varrho}_N^{-1}
\right) \leq L\dd_{FM,p}(P,Q)<+\infty.
\edeqn
\end{definition}

In comparison with the qualitative statistical robustness introduced by Kr\"atchmer et al. \cite{kratschmer2014comparative} or Cont et al. \cite{cont2010robustness}, the definition (\ref{eq:def-QSR-FM}) here has several advantages. 
First, we use Kantorovich metric instead of Prokhorov metric to quantify the discrepancy 
between $P^{\otimes N}\circ \widehat{\varrho}_N^{-1}$ and $Q^{\otimes N}\circ \widehat{\varrho}_N^{-1}$.
This enables us to capture the tail behaviour of the two laws
and facilitate us to derive an explicit bound for the difference. 
Second, we use the Fortet-Mourier metric to quantify the perturbation of  $P$, 
which is more sensitive than the L\'evy metric used in \cite{cont2010robustness} and 
the weighted Kolmogorov metric in \cite{kratschmer2014comparative,zahle2014qualitative,zahle2015qualitative,zahle2016definition} 
to the variation of the tails.
Third, inequality (\ref{eq:def-QSR-FM}) gives an error bound for the discrepancy of the two laws
and the bound is valid 
for all $Q$ in $\mathscr{M}$ instead of those in a neighborhood of $P$.

Next, we introduce a definition on the Lipschitz continuity of a general statistical mapping from $\mathscr{P}(\R)$ to $\R$, which strengthens the earlier definition of $\mathscr{C}$-continuity for a general statistical functional.

\begin{definition}
[Lipschitz continuity]
Let $\varrho:\mathscr{P}(\R)\to\R$ be a general statistical functional and $\mathscr{M}$ be a subset of $\mathscr{P}(\R)$. 
$\varrho$ is said to be \emph{Lipschitz continuous} on $\mathscr{M}$ w.r.t. $\dd$ if there exists a 
positive constant $L$ such that 
\bgeqn
\label{Lipschitz-continuous-condition}
|\varrho(P)-\varrho(Q)|\leq L\dd(P,Q)<+\infty,\quad \forall\; P,Q\in \mathscr{M}.
\edeqn
\end{definition}

There are a few of points to note to the above definition of Lipschitz continuity:
\begin{itemize}
  
\item[1.] 
The Lipschitz continuity is global instead of local over $\mathscr{M}$.
The condition is strong but we will find that many risk functionals are global Lipschitz continuous on some $\mathscr{M}$ indeed.

\item[2.] The magnitude of the continuity depends on the metric $\dd$ which measures the distance between $P$ and $Q$. In a specific case when $\dd=\dd_{FM,p}$, (\ref{Lipschitz-continuous-condition}) reduces to
\bgeqn
\label{eq:locally lipschitz continuous}
|\varrho(P)-\varrho(Q)|\leq L \int_{\R} |P(x)-Q(x)|c_p(x)dx<+\infty, \;\forall\;P,Q\in \mathscr{M},
\edeqn
where $c_p(x)=\max\{1,|x|^{p-1}\}$. The exponent $p$ plays an important role in (\ref{eq:locally lipschitz continuous}) because it interacts with the tails of $P(\cdot)$ and $Q(\cdot)$. 
Moreover, if $\mathcal{M}\subset\mathscr{P}_p(\R)$, then (\ref{eq:locally lipschitz continuous}) is finite. 
We will come back to this later. 

\item[3.] Let $P_N$ and $Q_N$ be empirical
distributions on $\R$. 
By plugging $P_N$ and $Q_N$ into (\ref{eq:locally lipschitz continuous}), we obtain
\bgeqn
\label{eq:locally lipschitz continuous-iid-one}
|\varrho(P_N)-\varrho(Q_N)| \notag
&\leq & L \int_{\R} |P_N(x)-Q_N(x)|c_p(x)dx
\\ \notag
&= & L \sum_{k=1}^{2N} \left|\frac{1}{N}\sum_{i=1}^N\mathbf{1}_{\xi^i\leq x_k}-\frac{1}{N}\sum_{i=1}^N\mathbf{1}_{\widehat{\xi}^i\leq x_k} \right| \int_{x_k}^{x_{k+1}} c_p(x)dx
\\ \notag
&= & L \sum_{k=1}^N \frac{1}{N} \left|\int_{\xi^{i_k}}^{\widehat{\xi}^{j_k}} c_p(x)dx \right|
\\ \notag
&\leq & L \sum_{k=1}^N \frac{1}{N} |\xi^{i_k}-\widehat{\xi}^{j_k}|\max\{c_p(\xi^{i_k}),c_p(\widehat{\xi}^{j_k})\} 
\\ 
&\leq &
\frac{L}{N}\sum_{k=1}^N c_{p}(\xi^k,\widehat{\xi}^k)|\xi^k-\widehat{\xi}^k|,\quad \forall \xi^k,\widehat{\xi}^k\in \R,
\edeqn
where $x_k$ is the $k$-th smallest number among $\{\xi^1,\ldots,\xi^N;\widehat{\xi}^1,\ldots,\widehat{\xi}^N\}$ for $k=1,\ldots,2N$ and $x_{2N+1}=x_{2N}$ and  $c_p(\xi,\widehat{\xi})=\max\{1,|\xi|,|\widehat{\xi}|\}^{p-1}$ for all $\xi,\widehat{\xi}\in \R$.
The equality is due to Fubini's theorem for discrete case and the last inequality from Lemma \ref{lem:order-sum-ineq} for the non-decreasing sequences $\{\xi^{i_k}\max\{c_p(\xi^{i_k}),c_p(\widehat{\xi}^{j_k})\}\}_{k=1}^N$ and $\{\widehat{\xi}^{j_k}\max\{c_p(\xi^{i_k}),c_p(\widehat{\xi}^{j_k})\}\}_{k=1}^N$.

{\color{black}
\item[4.]
In the case when $\varrho$ is continuous on $\mathscr{M}$, the Lipschitz continuity (\ref{eq:locally lipschitz continuous}) is equivalent to the Lipschitz continuity (\ref{eq:locally lipschitz continuous-iid-one}) on the set of all empirical distributions $\mathscr{M}_{1,\mathrm{emp}}$  
(see the first inequality of equation (\ref{eq:locally lipschitz continuous-iid-one})) 
because $\mathscr{M}_{1,\mathrm{emp}}$ is dense in $\mathscr{P}(\R)$.
}

\end{itemize}

\begin{example}[$p$-th moment functional]
For $p\geq 1$, we consider the \emph{$p$-th moment functional} $T^{(p)}$ on $\mathcal{M}_1^p=\mathscr{P}_p(\R)$ as defined by: 
\[
T^{(p)}(P):=\int_{-\infty}^{+\infty} x^p dP(x)<+\infty,\quad \forall P\in \mathcal{M}_1^p.
\]
Analogous to Example \ref{ex:counter-example}, we have 
\bgeqn
\label{ex:p-th-moment}
T^{(p)}(P)=-\int_{-\infty}^0 P(x)px^{p-1}dx+ \int_{0}^{+\infty} (1-P(x))px^{p-1}dx.
\edeqn
Thus, for any $P,Q\in \mathcal{M}_1^p$,
\bgeqn
\label{eq:p-th moment-local-Lip}
|T^{(p)}(P)-T^{(p)}(Q)| \notag &=&\left|\int_{-\infty}^{+\infty} (P(x)-Q(x))px^{p-1}dx\right|\leq p \int_{-\infty}^{+\infty} |P(x)-Q(x)||x|^{p-1}dx\\
&\leq& p \int_{-\infty}^{+\infty} |P(x)-Q(x)|c_p(x)dx<+\infty,
\edeqn
where $c_p(x)=\max\{1,|x|^{p-1}\}$. From (\ref{eq:locally lipschitz continuous}), we can see that the $p$-th moment functional $T^{(p)}$  is Lipschitz continuous w.r.t. $\dd_{FM,p}$  on $\mathcal{M}_1^p$.
\end{example}

\begin{lemma}
\label{L-Kant-estimate}
Let $\boldsymbol{\xi}:=(\xi^1,\cdots,\xi^N) \in \R^N$ and $\Psi$ be a set of functions from $\R^N$ to $\R$, i.e., 
\bgeqn
\label{lem:defined-set}
\Psi:=\left\{\psi: \R^N \to \R:
|\psi(\tilde{\boldsymbol{\xi}})-\psi(\widehat{\boldsymbol{\xi}})|\leq \frac{1}{N}\sum_{k=1}^N c_{p}(\tilde{\xi}^k,\widehat{\xi}^k)|\tilde{\xi}^k-\widehat{\xi}^k|,\;\forall \boldsymbol{\xi},\tilde{\boldsymbol{\xi}}\in \R^N\right\},
\edeqn
where $c_{p}(\xi,\tilde{\xi}):=\max\{1,|\xi|,|\tilde{\xi}|\}^{p-1}$ for all $\xi,\tilde{\xi}\in \R$ and $p\geq 1$. Then 
\bgeqn
\dd_{\Psi} (P^{\otimes N},Q^{\otimes N}) \leq \dd_{FM,p} (P,Q)<+\infty,\quad \forall P,Q\in \mathscr{P}_p(\R),
\edeqn
where $\dd_{\Psi}$ is defined by (\ref{def:zeta-metrics}).
\end{lemma}

Before presenting a proof, it might be helpful for us to explain why we consider a specific set of functions $\Psi$.
For fixed $N\in \mathbb{N}$, let $\mathscr{M}_{1,\mathrm{emp}}^N$ denote the set of
all empirical laws $P_{N}$ over $\R$, then $\mathscr{M}_{1,\mathrm{emp}}=\bigcup_{N\in \mathbb{N}}\mathscr{M}_{1,\mathrm{emp}}^N$. 
Then $\Psi$  may be regarded as a set of functions 
derived from a class of Lipschitz continuous functional 
on $\mathscr{M}_{1,\mathrm{emp}}^N$ with  $L=1$ and $\dd=\dd_{FM,p}$ (by writing $T(P_N)$ as a function of samples).
Lemma \ref{L-Kant-estimate} says that for any $N\in \mathbb{N}$, 
the discrepancy between $P^{\otimes N}$ and $Q^{\otimes N}$ under the metric $\dd_{\Psi}$ can be bounded by $\dd_{FM,p} (P,Q)$.

\noindent\textbf{Proof.}
Let $\xi^{-j}:=\{\xi^1,\cdots,\xi^{j-1},\xi^{j+1},\cdots,\xi^N\}$,
$\vec{\xi}^j:=\{\xi^1,\cdots,\xi^j\}$ and
$\vec{\xi}^{-j}:=\{\xi^{j+1},\cdots,\xi^N\}$.
For any $P_1,\cdots,P_N \in \mathscr{P}(\R)$ and any $j\in \{1,\cdots,N\}$, denote
\[
P_{-j}(d \xi^{-j}):=P_1(d \xi^1)\cdots P_{j-1}(d \xi^{j-1})P_{j+1}(d \xi^{j+1})\cdots P_N(d \xi^N)
\]
and
\[
h_{{\xi}^{-j}}({\xi}^j):=\int_{\R^{(N-1)}}
\psi({\xi}^{-j},\xi^j) P_{-j}(d \xi^{-j}).
\]
Then
\bgeq
|h_{{\xi}^{-j}}(\tilde{\xi}^j)-h_{{\xi}^{-j}}(\widehat{\xi}^j)|
&\leq& \int_{\R^{(N-1)}}
\left|\psi({\xi}^{-j},\tilde{\xi}^j)-\psi({\xi}^{-j},\widehat{\xi}^j)\right|
P_{-j}(d \xi^{-j})\\
&\leq& \int_{\R^{(N-1)}}
\frac{1}{N}c_{p}(\tilde{\xi}^j,\widehat{\xi}^j)|\tilde{\xi}^j-\widehat{\xi}^j|P_{-j}(d \xi^{-j})\\
&\leq& \frac{1}{N}c_{p}(\tilde{\xi}^j,\widehat{\xi}^j)|\tilde{\xi}^j-\widehat{\xi}^j|.
\edeq
Let $\mathcal{H}$ denote the set of functions $h_{{\xi}^{-j}}({\xi}^j)$ generated by $\psi\in \Psi$.
By the definition of $\dd_{\Psi}$ and the $p$-th order Forter-Mourier metric,
\bgeqn
\dd_{\Psi} (P_{-j} \times \tilde{P}_j,P_{-j} \times \widehat{P}_j)
&=&\sup_{\psi \in \Psi}
\left|
\int_{\R}\int_{\R^{(N-1)}}
\psi({\xi}^{-j},\xi^j) P_{-j}(d \xi^{-j}) \tilde{P}_j(d \xi^j)\right.\nonumber\\
&&-\left.
\int_{\R}\int_{\R^{(N-1)}}
\psi({\xi}^{-j},\xi^j) P_{-j}(d \xi^{-j}) \widehat{P}_j(d \xi^j)\right|\nonumber\\
&=& \sup_{h_{{\xi}^{-j}} \in {\cal H}} \left|
\int_{\R}h_{{\xi}^{-j}}({\xi}^j)\tilde{P}_j(d \xi^j)
-\int_{\R}h_{{\xi}^{-j}}({\xi}^j)\widehat{P}_j(d \xi^j) \right|\nonumber\\
&\leq& \frac{1}{N}\dd_{FM,p}(\tilde{P}_j,\widehat{P}_j),
\edeqn
where the inequality is due to $N h_{\xi^{-j}}(\xi^j)\in \mathcal{F}_{p}(\R)$ and the definition of $\dd_{FM,p}(P,Q)$. 
Finally, by the triangle inequality of the pseudo-metric, we have
\bgeq
\dd_{\Psi} \left(P^{\otimes N},
Q^{\otimes N}\right) 
&\leq&\dd_{\Psi} \left(P^{\otimes N}, P^{\otimes (N-1)}\times Q\right)+ \dd_{\Psi} \left(P^{\otimes (N-1)}\times Q, P^{\otimes (N-2)}\times Q^{\otimes 2}\right)\\
&&+ \cdots+ \dd_{\Psi} \left(P\times Q^{\otimes (N-1)}, Q^{\otimes N} \right)\\
&\leq& \frac{1}{N} \dd_{FM,p}(P,Q) \times N\\
&=& \dd_{FM,p}(P,Q).
\edeq
The proof is complete.
\hfill $\Box$

With the intermediate technical result, 
we are now ready to present our main result of quantitative statistical robustness 
for the plug-in estimator of a general risk functional.

\begin{theorem}
\label{thm:quant-stat-robust-T}
Let $\varrho:\mathscr{P}(\R)\to\R$ be a general statistical functional and $\mathscr{M}$ be a subset of $\mathscr{P}_p(\R)$ with $p\geq 1$. 
Assume, for fixed $N\in \mathbb{N}$, there exists a positive constant $L$ such that  
\bgeqn
|\varrho(P_N)-\varrho(Q_N)| \leq \frac{L}{N}\sum_{k=1}^N c_{p}(\xi^k,\widehat{\xi}^k)|\xi^k-\widehat{\xi}^k|,\;\forall \xi^k,\widehat{\xi^k}\in \R,
\label{eq:T_N-Lip}
\edeqn
where $P_N$ and $Q_N$ are given by (\ref{eq:emp-prob-P-N}) and (\ref{eq:emp-prob-Q-N}) respectively. 
Then $\widehat{\varrho}_N$
is quantitatively robust on $\mathscr{M}$ w.r.t. $(\dd_{K},\dd_{FM,p})$, i.e.,
\bgeqn
\dd_{K}\left(P^{\otimes N}\circ \widehat{\varrho}_N^{-1},
Q^{\otimes N}\circ \widehat{\varrho}_N^{-1}\right) \leq L\dd_{FM,p}(P,Q)<+\infty,\;\forall\; P, Q\in \mathscr{M}.
\label{eq:quant-SR-zeta-metric-FM}
\edeqn
If (\ref{eq:T_N-Lip}) holds for  all $N\in\mathbb{N}$, then the whole sequence of the plug-in estimators $\{\widehat{\varrho}_N\}_{N\in \mathbb{N}}$ is quantitatively robust on $\mathscr{M}$, i.e., (\ref{eq:quant-SR-zeta-metric-FM}) holds for all $N\in\mathbb{N}$.
Moreover, in the case when $p=1$, (\ref{eq:quant-SR-zeta-metric-FM}) reduces to
\bgeqn
\dd_{K} \left(P^{\otimes N}\circ \widehat{\varrho}_N^{-1},
Q^{\otimes N}\circ \widehat{\varrho}_N^{-1}\right) \leq L\dd_{K}(P,Q)<+\infty,\;\forall\; P,Q\in \mathscr{M}.
\label{eq:quant-SR-zeta-metric-K}
\edeqn
\end{theorem}

\noindent\textbf{Proof.}
Since the underlying probability space is atomless, then for any $N\in \mathbb{N}$, by definition
\bgeqn
&& \dd_{K} \left(P^{\otimes N}\circ \widehat{\varrho}_N^{-1},
Q^{\otimes N}\circ \widehat{\varrho}_N^{-1}\right) \nonumber\\
&=&\sup_{\psi \in \mathcal{F}_1(\R)}\left|
\int_\R \psi(t) P^{\otimes N}\circ \widehat{\varrho}_N^{-1}(dt) -
\int_\R \psi(t) Q^{\otimes N}\circ \widehat{\varrho}_N^{-1}(dt)
\right|\nonumber\\
&=&
\sup_{\psi \in \mathcal{F}_1(\R)}
\left|\int_{\R^N}\psi(\varrho(\vec{\xi}^N)) P^{\otimes N}(d\vec{\xi}^N)-\int_{\R^N} \psi(\varrho(\vec{\xi}^N)) Q^{\otimes N}(d\vec{\xi}^N)\right|,
\label{eq:SR-vt_n-thm-pseudo}
\edeqn
where we write $\vec{\xi}^N$ for $(\xi^1,\cdots,\xi^N)$ and $\varrho(\vec{\xi}^N)$ for $\widehat{\varrho}_N$ to indicate its dependence on $\xi^1,\cdots,\xi^N$. 

For any $\psi\in \mathcal{F}_1(\R)$, (\ref{eq:T_N-Lip}) ensures that
\bgeq
|\psi(\varrho(\tilde{\vec{\xi}}))-\psi(\varrho(\widehat{\vec{\xi}}))|\leq  |\varrho(\tilde{\vec{\xi}})-\varrho(\widehat{\vec{\xi}})| \leq \frac{L}{N} \sum_{k=1}^N c_{p}(\tilde{\xi}^k,\widehat{\xi}^k)|\tilde{\xi}^k-\widehat{\xi}^k|,\;\forall \tilde{\xi},\widehat{\xi}\in \R,
\edeq
which means that $\psi(\varrho(\cdot))$ is locally Lipschitz continuous in $\vec{\xi}^N$, i.e., $\psi(\varrho(\cdot))\in \mathcal{F}_p((\R^N)$ from (\ref{eq:locally lipschitz set}). 
Since $P, Q\in \mathscr{M}\subset \mathscr{P}_p(\R)\subset \mathscr{P}_{K}(\R)$ (see Example \ref{ex:admissible-law-probability-metrics}(i) and Proposition \ref{prop:class-FM}(i)), then (\ref{eq:SR-vt_n-thm-pseudo}) is finite. 
The rest follows from Lemma \ref{L-Kant-estimate} by
setting $\psi(\xi^1,\cdots,\xi^N)=\psi(\varrho(\xi^1,\cdots,\xi^N))$.

\hfill $\Box$


From Example \ref{ex:probability-metrics}, we have $\dd_{\mathrm{Prok}}(P,Q)\leq \sqrt{\dd_{K}(P,Q)}$ for all $P,Q\in \mathscr{P}(\R)$, then we have the following corollary. 
\begin{corollary}
Let $\varrho:\mathscr{P}(\R)\to\R$ be a general statistical functional. Assume that $\varrho$ is Lipschitz continuous w.r.t. $\dd_{FM,p}$ ($p\geq 1$) on $\mathscr{M}\subset \mathscr{P}_p(\R)$ for the constant $L$. 
Then the plug-in estimator sequence $\{\widehat{\varrho}_N\}_{N\in \mathbb{N}}$ is quantitatively robust on $\mathscr{M}$ w.r.t. $(\dd_{\mathrm{Prok}},\dd_{FM,p})$, i.e., 
\bgeq
\dd_{\mathrm{Prok}}\left(P^{\otimes N}\circ \widehat{\varrho}_N^{-1},
Q^{\otimes N}\circ \widehat{\varrho}_N^{-1}\right) \leq \sqrt{L\dd_{FM,p}(P,Q)}<+\infty,\;\forall\; P, Q\in \mathscr{M}
\edeq
for all $N\in\mathbb{N}$.
\end{corollary}

Next, we take a step further to consider the index of quantitative robustness for a general statistical functional. 

\begin{definition}[Index of quantitative robustness]
\label{def:index of quantitative robustness}
Let $\varrho:\mathscr{P}(\R)\rightarrow \R$ be a general statistical functional. 
If $\varrho$ is Lipschitz continuous w.r.t. $\dd_{FM,p}$ on $\mathscr{P}_p(\R)$ for the constant $L$ for some $p\geq 1$, then we can define an \emph{index of quantitative robustness of a statistical functional} $\varrho$ as
\bgeqn
\mathrm{iqr}(\varrho):=\left(\inf\{p\in [1,+\infty): \inmat{$\varrho$ is Lipschitz continuous w.r.t. $\dd_{FM,p}$}\; \inmat{on}\; \mathscr{P}_p(\R)\} \right)^{-1}.
\edeqn
\end{definition}
This index is a quantitative measurement for
the degree of robustness of a statistical functional. A larger index reflects a higher degree of robustness. 
For a general statistical functional $\varrho$, (\ref{eq:locally lipschitz continuous}) may hold for uncountable many $p$, see e.g., the $2$-th moment functional $T^{(2)}$ satisfying (\ref{eq:locally lipschitz continuous}) for any $p\geq 2$ on $\mathscr{P}_p(\R)=\mathcal{M}_1^p$. 
From Definition \ref{def:index of quantitative robustness}, we conclude that the $p$-th moment functional $T^{(p)}$ has the index $\mathrm{iqr}(T^{(p)})=\frac{1}{p}$. 
{\color{black}
Definition \ref{def:index of quantitative robustness} coincides with the index of qualitative robustness proposed by Kr\"atschmer et al. \cite{kratschmer2012qualitative} when $\varrho$ is Lipschitz continuous w.r.t. $\dd_{FM,p}$ on $\mathscr{P}_p(\R)$. The main advantage of Definition \ref{def:index of quantitative robustness} is that it is easy to calculate and we will illustrate this in the next section.
}

\section{Application to risk measures}

As we discussed in Proposition \ref{eq:RM-represent}, law invariant risk measure of a random variable
can be represented as a composition of a risk functional and law of
the random variable. In practice, risk of a random variable is
often calculated with empirical data, this is because either
the true probability distribution is unknown or it might be prohibitively expensive
to calculate the risk 
of a random variable with the true probability distribution.
This raises a question as to whether the estimated risk measure
based on empirical data is reliable or not. 
In this section, we apply the quantitative robustness results established 
in Theorem \ref{thm:quant-stat-robust-T} to some well-known risk measures.
The next proposition synthesizes   Proposition \ref{eq:RM-represent} and Theorem \ref{thm:quant-stat-robust-T}.

\begin{proposition} 
Let $\rho(X)$ be a tail-dependent law invariant convex risk measure with representation (\ref{eq:RM-represent}), let $P_N$ and $Q_N$ be empirical probability measures defined as in (\ref{eq:emp-prob-Q-N}). Assume that there exists a positive number $p\geq 1$ such that 
\bgeqn
\label{eq:assumption-estimator-1}
|\varrho(P_N)-\varrho(Q_N)| \leq 
\frac{L}{N}\sum_{k=1}^N c_{p}(\xi^k,\widehat{\xi}^k)|\xi^k-\widehat{\xi}^k|,\forall \vec{\xi},\widehat{\vec{\xi}}\in\R^N.
\edeqn
Then for any $N \in \mathbb{N}$ and any $P, Q\in {\cal M}_{1}^{p}$
\bgeqn
\dd_{K} \left(P^{\otimes N}\circ \varrho(P_N)^{-1},
Q^{\otimes N}\circ \varrho(Q_N)^{-1}\right) \leq L\dd_{FM,p}(P,Q)<+\infty.
\label{eq:quant-SR-zeta-metric-FM-app}
\edeqn
\end{proposition}

In what follows, we verify condition (\ref{eq:assumption-estimator-1})
for some well-known risk measures and hence show that they 
satisfy the proposed quantitative statistical robustness (\ref{eq:quant-SR-zeta-metric-FM-app}). 
To make the notation easily, we introduce the law invariant risk measure on the space of probability distributions.

\begin{example}
	The \emph{expectation} of $G\in \mathscr{P}(\R)$ given by
	$
	\mathbb{E}(G):=\int_{\R} \xi dG(\xi)
	$
	satisfies 
	\bgeq
	|\mathbb{E}(P_N)-\mathbb{E}(Q_N)|=\left|\int_{\R}\xi d(P_N-Q_N)(\xi)\right|\leq \frac{1}{N}\sum_{i=1}^N|\xi^i-\widehat{\xi}^i|.
	\edeq
	Let $T_N:=\mathbb{E}(\widehat{G}_N)$, where $\widehat{G}_N$ is the empirical distribution of $G$. 
	Then for any $N \in \mathbb{N}$ and any $P, Q\in {\cal M}_{1}^{1}$, 
	\bgeqn	
	\dd_{K} \left(P^{\otimes N}\circ T_N^{-1},Q^{\otimes N}\circ T_N^{-1}\right) \leq \dd_{K}(P,Q)<+\infty,
	\edeqn
	and the index of quantitative robustness  $\mathrm{iqr}(\mathbb{E})=1$.
\end{example}

\begin{example}
Consider the \emph{conditional value-at-risk} of a probability distribution $G\in \mathscr{P}(\R)$ at level $\tau\in (0,1)$,
which is defined by
	\bgeq
	\inmat{CVaR}_\tau(G):=\inf \left\{r+\frac{1}{1-\tau}\int_{\R}\max\{0,\xi-r\} dG(\xi),\forall r\in\R\right\}.
	\edeq
Then
	\bgeq
	|\inmat{CVaR}_p(P_N)-\inmat{CVaR}_p(Q_N)|
	&\leq& \frac{1}{1-\tau}\sup_{r\in\R}\left|\int_{\R}\max\{0,\xi-r\}d(P_N-Q_N)(\xi) \right|\\
	&=&\frac{1}{1-\tau}\sup_{r\in\R}\frac{1}{N} \left|\sum_{i=1}^N\max\{0,\xi^i-r\}-\max\{0,\widehat{\xi}^i-r\}\right|\\
	&\leq& \frac{1}{1-\tau}\times\frac{1}{N}\sum_{i=1}^N|\xi^i-\widehat{\xi}^i|,
	\edeq	
	the last inequality is due to the fact that $|\max\{0,x\}-\max\{0,y\}|\leq |x-y|$ holds for all $x, y\in \R$.
	
	Let $T_N:=\inmat{CVaR}_{\tau}(\widehat{G}_N)$, where $\widehat{G}_N$ is the empirical distribution of $G$. 
	Then for any $N \in \mathbb{N}$ and any $P, Q\in {\cal M}_{1}^{1}$, 
	\bgeqn	
	\dd_{K} \left(P^{\otimes N}\circ T_N^{-1},Q^{\otimes N}\circ T_N^{-1}\right) \leq \frac{1}{1-\tau}\dd_{K}(P,Q)<+\infty, 
	\edeqn
	and the index of quantitative robustness  $\mathrm{iqr}(\inmat{CVaR}_\tau)=1$ for $\tau\in (0,1)$.
\end{example}

\begin{example}
	The \emph{upper semi-deviation} $sd_+(G)$ of a measure $G\in\mathscr{P}(\R)$, which is defined by
	\bgeq
	sd_+(G):=\int_{\R}\max\left\{0,\xi-\int_{\R}udG(u) \right\}dG(\xi),
	\edeq
	satisfies 
	\bgeq
	|sd_+(P_N)-sd_+(Q_N)|
	&= & \left|\frac{1}{N}\sum_{j=1}^N\max\left\{0, \xi^j-\frac{1}{N}\sum_{i=1}^N \xi^i \right\}-\frac{1}{N}\sum_{j=1}^N\max\left\{0, \widehat{\xi}^j-\frac{1}{N}\sum_{i=1}^N \widehat{\xi}^i \right\}\right|\\
	&\leq & \frac{1}{N}\sum_{j=1}^N\left|\max\left\{0, \xi^j-\frac{1}{N}\sum_{i=1}^N \xi^i \right\}-\max\left\{0, \widehat{\xi}^j-\frac{1}{N}\sum_{i=1}^N \widehat{\xi}^i \right\} \right|\\
	&\leq & \frac{1}{N}\sum_{j=1}^N \left|\left(\xi^j-\frac{1}{N}\sum_{i=1}^N \xi^i\right)- \left(\widehat{\xi}^j-\frac{1}{N}\sum_{i=1}^N \widehat{\xi}^i\right) \right|\\
	&\leq & \frac{1}{2}\sum_{j=1}^N \left( \left|\xi^j-\widehat{\xi}^j \right|+ \frac{1}{N}\sum_{i=1}^N|\xi^i-\widehat{\xi}^i| \right)\\
	&= & \frac{2}{N} \sum_{i=1}^N|\xi^i-\widehat{\xi}^i|.
	\edeq
	Let $T_N:=sd_+(\widehat{G}_N)$, where $\widehat{G}_N$ is the empirical distribution of $G$. 
	Then for any $N \in \mathbb{N}$ and any $P, Q\in {\cal M}_{1}^{1}$, 
	\bgeqn	
	\dd_{K} \left(P^{\otimes N}\circ T_N^{-1},Q^{\otimes N}\circ T_N^{-1}\right) \leq 2\dd_{K}(P,Q)<+\infty,
	\edeqn
	and the index of quantitative robustness  $\mathrm{iqr}(\inmat{sd}_+)=1$.
\end{example}

\begin{example}
The \emph{Optimized Certainty Equivalent (OCE)} \cite{ben2007old} of $G\in \mathscr{P}(\R)$ is given by
\bgeq
S_{u}(G):=\sup_{\eta\in\R}\left\{\eta+\int_{\R} u(\xi-\eta)dG(\xi)\right\},
\edeq
where $u:\R\rightarrow [-\infty,\infty)$ is a proper concave and non-decreasing utility function satisfying the normalized property: $u(0)=0$ and $1\in \partial u(0)$, where $\partial u(\cdot)$ denotes the subdifferential map of $u$.
By the essential of \cite[Proposition 2.1]{ben2007old}, we have 
\bgeq
S_{u}(P_N)=\sup_{\eta\in \R} \left\{\eta+\frac{1}{N}\sum_{i=1}^N u(\xi^i-\eta) \right\}
=\sup_{\eta\in [\xi_{\min},\xi_{\max}]} \left\{\eta+\frac{1}{N}\sum_{i=1}^N u(\xi^i-\eta) \right\},
\edeq
where $\xi_{\min}=\min\{\xi^1,\ldots,\xi^N;\widehat{\xi}^1,\ldots,\widehat{\xi}^N\}$ and $\xi_{\max}=\max\{\xi^1,\ldots,\xi^N; \widehat{\xi}^1,\ldots,\widehat{\xi}^N\}$. 
Let $\rho(G):=-	S_{u}(G)$. 
Then
$\rho(\cdot)$ is a convex risk measure \cite{ben2007old} and
\bgeq
|\rho(P_N)-\rho(Q_N)|
&\leq & \sup_{\eta\in [\xi_{\min},\xi_{\max}]}\left|\left(\eta+\int_{\R} u(\xi-\eta)dP_N(\xi)\right)-\left(\eta+\int_{\R} u(\xi-\eta)dQ_N(\xi)\right) \right|\\
&= & \sup_{\eta\in [\xi_{\min},\xi_{\max}]}\left| \frac{1}{N}\sum_{i=1}^N u(\xi^i-\eta)-\frac{1}{N}\sum_{i=1}^N u(\widehat{\xi}^i-\eta) \right|\\
&\leq & \sup_{\eta\in [\xi_{\min},\xi_{\max}]} \frac{1}{N}\sum_{i=1}^N\left|u(\xi^i-\eta)- u(\widehat{\xi}^i-\eta) \right|\\
&\leq & \frac{1}{N}\sum_{i=1}^N u'_-(\xi_{\min})
|\xi^i-\widehat{\xi}^i|,
\edeq
where $u'_-(t)$ denotes the left derivative of $u$ at $t$ and the last inequality is due to the fact that $u$ is non-decreasing and concave, subsequently, $u'_-(t)$ is non-increasing.

Let $T_N:=-S_u(\widehat{G}_N)$, where $\widehat{G}_N$ is the empirical distribution of $G$. We consider two interesting cases.

One is that 
$\sup_{\eta\in\R}u'_-(\eta)<+\infty$, in which case
\bgeqn	
\dd_{K} \left(P^{\otimes N}\circ T_N^{-1},Q^{\otimes N}\circ T_N^{-1}\right) \leq \sup_{\eta\in\R}u'_-(\eta)\dd_{K}(P,Q)<+\infty,
\label{eq:OCE-bounded}
\edeqn
for any $N \in \mathbb{N}$ and any $P, Q\in {\cal M}_{1}^{1}$ and the index of quantitative robustness for this case is $1$.

The other is that 
there exists some positive number $p>1$ and positive constant $L$ such that $u'_-(\xi_{\min})\leq L c_{p}(\xi^i,\widehat{\xi}^i)$, 
where $c_p(\xi^i,\widehat{\xi}^i)=\max\{1,|\xi^i|,|\widehat{\xi}^i|\}^{p-1}$.
In that case, we have
\bgeqn	
\dd_{K} \left(P^{\otimes N}\circ T_N^{-1},Q^{\otimes N}\circ T_N^{-1}\right) \leq L\dd_{FM,p}(P,Q)<+\infty,
\label{eq:OCE-local-Lip}
\edeqn
and the index of quantitative robustness for this case is $\frac{1}{p}$.

To see how (\ref{eq:OCE-bounded}) and (\ref{eq:OCE-local-Lip}) could possibly be satisfied, we consider two
specific utility functions: piecewise linear utility function and quadratic utility function, both of which are extracted from  \cite{ben2007old}.

(a)  Piecewise linear utility function 
with $u(t):=\gamma_1[t]_++\gamma_2[-t]_+$, where $0\leq \gamma_1 < 1 < \gamma_2$ and $[z]_+=\max\{0,z\}$.
A simple calculation yields
\bgeq
|\rho(P_N)-\rho(Q_N)|\leq \frac{\gamma_2}{N}\sum_{i=1}^N|\xi^i-\widehat{\xi}^i|.
\edeq
Thus for any $N \in \mathbb{N}$ and any $P, Q\in {\cal M}_{1}^1$, 
\bgeqn	
\dd_{K} \left(P^{\otimes N}\circ T_N^{-1},Q^{\otimes N}\circ T_N^{-1}\right) \leq \gamma_2\dd_{K}(P,Q)<+\infty,
\edeqn
and and the index of quantitative robustness $\mathrm{iqr}(-S_u)=1$.

(b) Quadratic utility 
with $u(t):=(t-\frac{1}{2}t^2)\mathbf{1}_{(-\infty,1)}(t)+\frac{1}{2}\mathbf{1}_{[1,+\infty)}(t)$.
It is easy to observe that the function is locally Lipschitz continuous over $[\xi_{\min},\xi_{\max}]$ with modulus being bounded by $|1-\xi_{\min}|$. Thus
\bgeq
|\rho(P_N)-\rho(Q_N)|&\leq& \sup_{\eta\in [\xi_{\min},\xi_{\max}]}\frac{1}{N} \sum_{i=1}^N\left|u(\xi^i-\eta)-u(\widehat{\xi}^i-\eta)\right|\\
&\leq& \frac{1}{N}\sum_{i=1}^N |1-\xi_{\min}||\xi^i-\widehat{\xi}^i|.
\edeq
Moreover, if $\xi_{\min}\leq -1$, then $|1-\xi_{\min}|\leq 2|\xi_{\min}|$. Subsequently, 
\[
|\rho(P_{N})-\rho(Q_{N})|\leq \frac{2}{N}\sum_{i=1}^N c_2(\xi^i,\widehat{\xi}^i)|\xi^i-\widehat{\xi}^i|,
\]
where $c_2(\xi^i,\widehat{\xi}^i)=\max\{1,|\xi^i|,|\widehat{\xi}^i|\}$. 
Thus for any $N \in \mathbb{N}$ and any $P, Q\in {\cal M}_{1}^2$, 
\bgeqn	
\dd_{K} \left(P^{\otimes N}\circ T_N^{-1},Q^{\otimes N}\circ T_N^{-1}\right) \leq 2\dd_{FM,2}(P,Q)
\edeqn
provided that $\xi_{\min}<-1$ and the index of quantitative robustness $\mathrm{iqr}(-S_u)=\frac{1}{2}$.
\end{example}

\begin{example}
Suppose that $l: \R \rightarrow\R$ is an increasing convex loss function which is not identically constant. 
Let $x_0$ be an interior point in the range of $l$.
The \emph{Shortfall Risk Measure} \cite{follmer2002convex} of $G\in \mathscr{P}(\R)$ is defined by 
\bgeqn
\rho_{l}(G):=\inf\left\{m\in \R: 
\int_{\R}l(\xi-m) dG(\xi)\leq x_0\right\}.
\label{eq:Shortfall-RM}
\edeqn
Following a similar analysis to Guo and Xu \cite{guo2017convergence}, we can recast the formulation above as 
\bgeqn
\rho_{l}(G)=\inf_{m\in \R}\sup_{\lambda\geq 0}\left\{m+\lambda\left(\int_{\R}l(\xi-m)dG(\xi)-x_0\right)\right\}.
\label{eq:Shortfall-RM-1}
\edeqn
Swapping the inf and sup operations, we can obtain the Lagrange dual of the problem. Moreover,
if we assume that the inequality constraint in (\ref{eq:Shortfall-RM}) satisfies the well-known Slater condition,
i.e., there exists $m_0$ such that $\int_\R l(\xi-m_0)dG(\xi)-x_0<0$, then the Lagrange multipliers of 
(\ref{eq:Shortfall-RM}) is bounded and the strong duality holds. Consequently, we can rewrite (\ref{eq:Shortfall-RM-1}) as
\bgeqn
\rho_{l}(G)=\inf_{m\in \R}\sup_{\lambda\in [a,b]}\left\{m+\lambda\left(\int_{\R}l(\xi-m)dG(\xi)-x_0\right)\right\},
\label{eq:Shortfall-RM-2}
\edeqn
where $a, b$ are some positive numbers.
By the essential of \cite[Proposition 2.1]{ben2007old}, we have 
\bgeq
\rho_{l}(P_N) &=& \sup_{\lambda\in [a,b]}\inf_{m \in \R} \left\{m+\lambda\left(\frac{1}{N}\sum_{i=1}^N l(\xi^i-\eta) -x_0\right)\right\}\\
&=&\sup_{\lambda\in [a,b]}\inf_{m\in [\xi_{\min},\xi_{\max}]} \left\{m+\lambda\left(\frac{1}{N}\sum_{i=1}^N l(\xi^i-m) -x_0\right)\right\},
\edeq
where $\xi_{\min}=\min\{\xi^1,\ldots,\xi^N;\widehat{\xi}^1,\ldots,\widehat{\xi}^N\}$ and $\xi_{\max}=\max\{\xi^1,\ldots,\xi^N; \widehat{\xi}^1,\ldots,\widehat{\xi}^N\}$. 
Subsequently, 
\bgeq
|\rho_{l}(P_N)-\rho_{l}(Q_N)|
&\leq& b\sup_{m\in [\xi_{\min},\xi_{\max}]} \left|\frac{1}{N}\sum_{i=1}^N l(\xi^i-m)- \frac{1}{N}\sum_{i=1}^N l(\widehat{\xi}^i-m)\right|\\
&\leq& b\sup_{m\in [\xi_{\min},\xi_{\max}]}\frac{1}{N} \sum_{i=1}^N\left| l(\xi^i-m)-l(\widehat{\xi}^i-m)\right|\\
&\leq & b\sup_{m\in [\xi_{\min},\xi_{\max}]} \frac{1}{N} \sum_{i=1}^N [l'_+(\xi^i-m)\vee l'_+(\widehat{\xi}^i-m)]|\xi^i-\widehat{\xi}^i|\\
&\leq & \frac{b}{N} \sum_{i=1}^N [l'_+(\xi^i-\xi_{\min})\vee l'_+(\widehat{\xi}^i-\xi_{\min})]|\xi^i-\widehat{\xi}^i| \\
&\leq & \frac{b}{N}\sum_{i=1}^N \sup_{m\in \R}l'_+(m)|\xi^i-\widehat{\xi}^i|,
\edeq
where $l'_+(t)$ denote the right derivative of $l$ at $t$ and the last three inequalities are due to the fact $l$ is non-decreasing convex, subsequently, $l'_+(t)$ is non-decreasing. 

Let $T_N:=\rho_{l}(\widehat{G}_N)$, where $\widehat{G}_N$ is the empirical distribution of $G$. 
If $\sup_{m\in \R} l'_+(m)<+\infty$, 
then for any $N \in \mathbb{N}$ and any $P, Q\in {\cal M}_{1}^{1}$, 
\bgeqn	
\dd_{K} \left(P^{\otimes N}\circ T_N^{-1},Q^{\otimes N}\circ T_N^{-1}\right) \leq \sup_{m\in\R}l'_+(\eta)\dd_{K}(P,Q)<+\infty.
\edeqn
If there exists some positive number $p>1$ and positive constant $L$ such that $l'_+(\xi^i-\xi_{\min})\vee l'_+(\widehat{\xi}^i-\xi_{\min})\leq L c_{p}(\xi^i,\widehat{\xi}^i)$, 
where $c_p(\xi^i,\widehat{\xi}^i)=\max\{1,|\xi^i|,|\widehat{\xi}^i|\}^{p-1}$, 
then
\bgeqn	
\dd_{K} \left(P^{\otimes N}\circ T_N^{-1},Q^{\otimes N}\circ T_N^{-1}\right) \leq L\dd_{FM,p}(P,Q)<+\infty.
\edeqn
In what follows, we illustrate the above two inequalities with two specific loss functions: deposit insurance loss function \cite{chen2013axiomatic} and $p$-th power loss function \cite{follmer2002convex}.

(a) Deposit insurance loss function, $l(x)=[x]_+$, where $[x]_+=\max\{x,0\}$. Then $\sup_{m\in \R} l'_+(m)<+\infty$. Thus, for any $N \in \mathbb{N}$ and any $P, Q\in {\cal M}_{1}^{\phi}$, 
\bgeqn	
\dd_{K} \left(P^{\otimes N}\circ T_N^{-1},Q^{\otimes N}\circ T_N^{-1}\right) \leq \sup_{m\in\R}l'_+(\eta)\dd_{K}(P,Q)<+\infty,
\edeqn
and the index of quantitative robustness is 1.

(b) For $x_0>0$, we consider the $p$-th power loss function, 
\[l(x)=
\begin{cases} 
\frac{1}{p} x^p, & \inmat{if}\;x \geq 0 \\
0,               & \inmat{otherwise}
\end{cases},
\]
where $p>1$. We have $l'_+(x)=x^{p-1}$ for $x\geq 0$ and $l'_+(x)=0$ for $x<0$. Then, if $\xi_{\min}\geq 0$, then $0\leq \xi^i-\xi_{\min}\leq |\xi^i|$ and subsequently 
$l'_+(\xi^i-\xi_{\min})\vee l'_+(\widehat{\xi}^i-\xi_{\min})\leq  c_{p}(\xi^i,\widehat{\xi}^i)$. Thus for any $N \in \mathbb{N}$ and any $P, Q\in {\cal M}_{1}^{p}$, 
\bgeqn	
\dd_{K} \left(P^{\otimes N}\circ T_N^{-1},Q^{\otimes N}\circ T_N^{-1}\right) \leq \dd_{FM,p}(P,Q)<+\infty
\edeqn
provided that $\xi_{\min}\geq 0$ and the index of quantitative robustness is $\frac{1}{p}$. 
\end{example}

In all of the above examples, 
the risk measures can either be represented explicitly in the form of $\int_\R f(x) dP(x)$ (such as Expectation)
or be obtained from solving an optimization problem where the underlying functions are represented in the expected utility form (CVaR, Certainty Equivalent and Shortfall risk measure), 
this is because the utility (disutility) functions are assumed to be concave (convex) and hence locally Lipschitz continuous. When growth of the Lipschitz  modulus  
is controlled by $c_p(\xi,\xi')$, these risk measures satisfy 
inequality (\ref{eq:locally lipschitz continuous}) as we have shown. 
This may not work for the spectral risk measures \cite{acerbi2002spectral}
with unbounded risk spectrum 
because the latter distort the probability distribution $P(x)$. 
However,
when the risk spectrum is bounded (such as CVaR which is a special case of spectral risk measure), 
we can still manage inequality  (\ref{eq:locally lipschitz continuous}). 
This explains why we haven't included spectral risk measures in the examples.



\bibliographystyle{ieeetr}
\bibliography{bibfile}


\begin{appendices}
\renewcommand\thefigure{\thesection.\arabic{figure}}
\renewcommand\thetable{\thesection.\arabic{table}}

\section{}

\setcounter{figure}{0}
\setcounter{table}{0}

\begin{lemma}
\label{lem:order-sum-ineq}
Let $\{a_i\}_{i=1}^N$ and $\{b_i\}_{i=1}^N$ be two non-decreasing sequences. Then for any permutation $\{k_1, k_2, \ldots, k_N\}$ of $\{1, 2, \ldots, N\}$, we have 
\bgeq
\sum_{i=1}^N |a_i-b_i|\leq \sum_{i=1}^N |a_i-b_{k_i}|.
\edeq
\end{lemma}

\noindent\textbf{Proof.} 
The result is perhaps well known. We include a proof as we cannot find a reference.
We do so by induction. 

For $N=1$, the statement is trivial and for $N=2$, $|a_1-b_1|+|a_2-b_2|\leq |a_1-b_2|+|a_2-b_1|$ for any $a_1\leq a_2$ and $b_1\leq b_2$. 
Assume that the conclusion holds for $N\leq n$.
Then for $N=n+1$, we have for any non-decreasing sequences $\{a_i\}_{i=1}^{n+1}$ and $\{b_i\}_{i=1}^{n+1}$ and any permutation of $\{k_1,\ldots,k_{n+1}\}$ of $\{1,\ldots,n+1\}$, 
there exists a $j\in\{1,\ldots,n+1\}$ such that $b_{k_j}=b_{n+1}$. 
If $j=n+1$, then from induction hypothesis for $N=n$, we have 
\bgeq
\sum_{i=1}^{n+1} |a_i-b_i|=\sum_{i=1}^{n} |a_i-b_i|+|a_{n+1}-b_{k_{n+1}}|\leq \sum_{i=1}^{n+1} |a_i-b_{k_i}|. 
\edeq
If $j<n+1$, then we have 
\bgeq
\sum_{i=1}^{n+1}|a_i-b_{k_i}| &=& \sum_{i=1}^{j-1}|a_i-b_{k_i}|+\sum_{i=j+1}^{n}|a_i-b_{k_i}|+|a_j-b_{k_j}|+|a_{n+1}-b_{k_{n+1}}|\\
&\geq& \sum_{i=1}^{j-1}|a_i-b_{k_i}|+\sum_{i=j+1}^{n}|a_i-b_{k_i}|+|a_j-b_{k_{n+1}}|+|a_{n+1}-b_{k_j}|\\
&=&\sum_{i=1}^{j-1}|a_i-b_{k_i}|+\sum_{i=j+1}^{n}|a_i-b_{k_i}|+|a_j-b_{k_{n+1}}|+|a_{n+1}-b_{n+1}|\\
&\geq& \sum_{i=1}^{n}|a_i-b_i|+|a_{n+1}-b_{n+1}|=\sum_{i=1}^{n+1}|a_i-b_i|, 
\edeq
where the first inequality is from induction hypothesis for $N=2$ to the non-decreasing sequences $\{a_j,a_{n+1}\}$ and $\{b_{k_{n+1}},b_{k_j}\}$
and the second inequality is due to induction hypothesis for $N=n$ to the non-decreasing sequences $\{a_1,\ldots,a_n\}$ and $\{b_1,\ldots,b_n\}$.
\hfill $\Box$

\begin{proposition}
\label{prop:orderproduct}
Let $\{a_i\}_{i=1}^N$ be a sequence of numbers and $\{b_i\}_{i=1}^N$ be a sequence of non-negative numbers. If $a_{i_1}\leq a_{i_2}\leq \cdots \leq a_{i_N}$, $b_{i_1}\leq b_{i_2}\leq \cdots \leq b_{i_N}$, then
\bgeq
\sum_{k=1}^Na_kb_k\leq \sum_{k=1}^Na_{i_k}b_{i_k}.
\edeq
\end{proposition}
See e.g. \cite[Proposition 12]{kusuoka2001law}.

\section{}

\begin{example}
\label{ex:counter-example}
In this example, we show that both inclusions in (\ref{eq:relation_admissible_law_weighted_K}) are strict.
We first show that $\mathcal{M}_1^{\phi}\neq \mathscr{P}_{(\phi)}(\R)$, i.e., there exists a $P\in \mathscr{P}_{(\phi)}(\R)$ such that $P\notin \mathcal{M}_1^{\phi}$. 
Let $\phi$ be a unbounded $u$-shaped function. Then by the continuity of $\phi$, there exist $a<0$ and $b>0$ with $\phi(a)=2=\phi(b)$. 
Let 
\bgeq
P(x)=
\begin{cases} 
\frac{1}{\phi(x)}, & \inmat{for}\;x \leq a, \\
\frac{1}{2}, & \inmat{for}\; a\leq x\leq b\\
1-\frac{1}{\phi(x)},&\inmat{for} \; x\geq b.
\end{cases},
\edeq
Since $\phi(x)\geq 1$ for all $x$ outside $[a,b]$, then $P(x)$ is well-defined on $\R$. 
By the monotonicity and unboundedness of $\phi$, we have $P\in \mathscr{P}(\R)$. 
Moreover, since 
$$
\sup_{x\leq 0}|P(x)\phi(x)|+\sup_{x>0}|(1-P(x))\phi(x)|=2,
$$ 
then $P\in \mathscr{P}_{(\phi)}(\R)$. However, by change of variables in integration, we have
\bgeq
\int_{\R} \phi(x)dP(x) 
&=& \int_{-\infty}^a \phi(x)d\left(\frac{1}{\phi(x)}\right)+\int_{b}^{+\infty} \phi(x)d\left(1-\frac{1}{\phi(x)}\right)\\
&=& \int_0^{\frac{1}{2}}\frac{1}{t}dt + \int_{\frac{1}{2}}^1 \frac{1}{1-t} dt= 2\int_0^{\frac{1}{2}}\frac{1}{t}dt
= +\infty,
\edeq
which means $P\notin \mathcal{M}_{1}^{\phi}$.

Now we show that $\mathscr{P}_{(\phi)}(\R)\neq \bigcap_{\epsilon>0} \mathcal{M}_1^{\phi^{1-\epsilon}}$, i.e., there exists a $P\in \bigcap_{\epsilon>0} \mathcal{M}_1^{\phi^{1-\epsilon}}$ such that $P\notin \mathscr{P}_{(\phi)}(\R)$.
Let $\phi$ be an unbounded $u$-shaped function. Then there exists an unbounded $u$-shaped function $\psi$ such that $\lim_{|x|\rightarrow +\infty}\psi(x)/\phi(x)=0$. 
More precisely, for any $\epsilon \in (0,1)$, there exists an unbounded $u$-shaped function $\psi$ such that 
\bgeqn
\label{phi-growth-condition}
\lim_{|x|\rightarrow +\infty}\psi(x)/\phi(x)^{1-\epsilon}=0.
\edeqn
We construct such $\psi$ as follows: 
since $\phi$ is an unbounded $u$-shape function, then there exist $a<0$ and $b>0$ with $\phi(a)=e^2=\phi(b)$. 
Let
\bgeq
\psi(x)=
\begin{cases}
\ln{(\phi(x))}, & \inmat{for}\;x \leq a, \\
2, & \inmat{for}\; a\leq x\leq b,\\
\ln{(\phi(x))}, &\inmat{for} \; x\geq b.
\end{cases}
\edeq
Then $\psi$ is an unbounded $u$-shaped function and satisfies (\ref{phi-growth-condition}). 
Let 
\bgeq
P(x)=
\begin{cases} 
\frac{1}{\psi(x)}, & \inmat{for}\;x \leq a, \\
\frac{1}{2}, & \inmat{for}\; a\leq x\leq b\\
1-\frac{1}{\psi(x)},&\inmat{for} \; x\geq b.
\end{cases},
\edeq
Since $\psi(x)\geq 1$ for all $x$, then $P(x)$ is well-defined on $\R$. 
By the monotonicity and unboundedness of $\psi$, we have $P\in \mathscr{P}(\R)$.  

For fixed $\epsilon\in (0,1)$, by change of variables in integration, we have 
\bgeq
\int_{\R}\phi(x)^{1-\epsilon}dP(x)
&=&\int_{-\infty}^a \phi(x)^{1-\epsilon} d\left(\frac{1}{\psi(x)}\right)+\int_{b}^{+\infty} \phi(x)^{1-\epsilon}d \left(1-\frac{1}{\psi(x)}\right)\\
&=& \int_{-\infty}^a \phi(x)^{1-\epsilon} d\left(\frac{1}{\ln{\phi(x)}}\right)+\int_{b}^{+\infty} \phi(x)^{1-\epsilon}d \left(1-\frac{1}{\ln{\phi(x)}}\right)\\
&=& \int_0^{\frac{1}{\ln{2}}} e^{\frac{1-\epsilon}{t}} dt+ \int_{\frac{1}{\ln{2}}}^1 e^{\frac{1-\epsilon}{1-t}}dt \\
&<& +\infty.
\edeq
Since for $\epsilon\geq 1$, $\phi^{1-\epsilon}$ is bounded on $\R$, then $\mathcal{M}_{1}^{\phi^{1-\epsilon}}=\mathscr{P}(\R)$. 
Thus, $P\in \bigcap_{\epsilon>0} \mathcal{M}_1^{\phi^{1-\epsilon}}$. 
However, 
\bgeq
\sup_{x\leq 0}|P(x)\phi(x)|+\sup_{x>0}|(1-P(x))\phi(x)|
\geq\sup_{x\leq a} \left|\frac{\phi(x)}{\ln{(\phi(x))}}\right|+\sup_{x\geq b}\left|\frac{\phi(x)}{\ln{(\phi(x))}}\right|=+\infty,
\edeq
which means $P\notin \mathscr{P}_{(\phi)}(\R)$.

%
%

\end{example}

\end{appendices}

\end{document}